\documentclass[sigconf]{acmart}

\AtBeginDocument{%
  \providecommand\BibTeX{{%
    \normalfont B\kern-0.5em{\scshape i\kern-0.25em b}\kern-0.8em\TeX}}}

\copyrightyear{2025}
\acmYear{2025}
\setcopyright{acmlicensed}\acmConference[ASSETS '25]{The 27th International ACM SIGACCESS Conference on Computers and Accessibility}{October 26--29, 2025}{Denver, CO, USA}
\acmBooktitle{The 27th International ACM SIGACCESS Conference on Computers and Accessibility (ASSETS '25), October 26--29, 2025, Denver, CO, USA}
\acmDOI{10.1145/3663547.3746320}
\acmISBN{979-8-4007-0676-9/2025/10}
\setcopyright{acmlicensed}

\usepackage{subfigure}
\usepackage{subcaption}
\usepackage{tcolorbox}
\usepackage{tabularx}

\newcommand{\softwarename}{\emph{DescribePro}}
\usepackage[utf8]{inputenc}
\usepackage{enumitem} 
\definecolor{lightgray}{gray}{0.9}

\usepackage{xcolor}

\definecolor{lightblue}{RGB}{244, 246, 253}
\begin{document}

\title[DescribePro: Collaborative Audio Description with Human-AI Interaction]{DescribePro: Collaborative Audio Description with Human-AI Interaction}

\author{Maryam Cheema}
\affiliation{
  \institution{Arizona State University}
  \city{Tempe}
  \state{Arizona}
  \country{USA}
}
\email{mcheema2@asu.edu}

\author{Sina Elahimanesh}
\affiliation{%
  \institution{Sharif University of Technology}
  \city{Tehran}
  \country{Iran}
}
\email{sina.elahimanesh@sharif.edu}

\author{Samuel Martin}
\affiliation{
  \institution{Arizona State University}
  \city{Tempe}
  \state{Arizona}
  \country{USA}
}
\email{simart12@asu.edu}

\author{Pooyan Fazli}
\affiliation{
  \institution{Arizona State University}
  \city{Tempe}
  \state{Arizona}
  \country{USA}
}
\email{pooyan@asu.edu}

\author{Hasti Seifi}
\affiliation{
  \institution{Arizona State University}
  \city{Tempe}
  \state{Arizona}
  \country{USA}
}
\email{hasti.seifi@asu.edu}


\begin{abstract}
Audio description (AD) makes video content accessible to millions of blind and low vision (BLV) users. However, creating high-quality AD involves a trade-off between the precision of human-crafted descriptions and the efficiency of AI-generated ones. To address this, we present \softwarename\, a collaborative AD authoring system that enables describers to iteratively refine AI-generated descriptions through multimodal large language model prompting and manual editing. \softwarename\ also supports community collaboration by allowing users to fork and edit existing ADs, enabling the exploration of different narrative styles.
We evaluate \softwarename\ with 18 describers (9 professionals and 9 novices) using quantitative and qualitative methods. 
Results show that AI support reduces repetitive work while helping professionals preserve their stylistic choices and easing the cognitive load for novices. Collaborative tags and variations show potential for providing customizations, version control, and training new describers. 
These findings highlight the potential of collaborative, AI-assisted tools to enhance and scale AD authorship.
\end{abstract}




\begin{teaserfigure}
  \includegraphics[width=\textwidth]{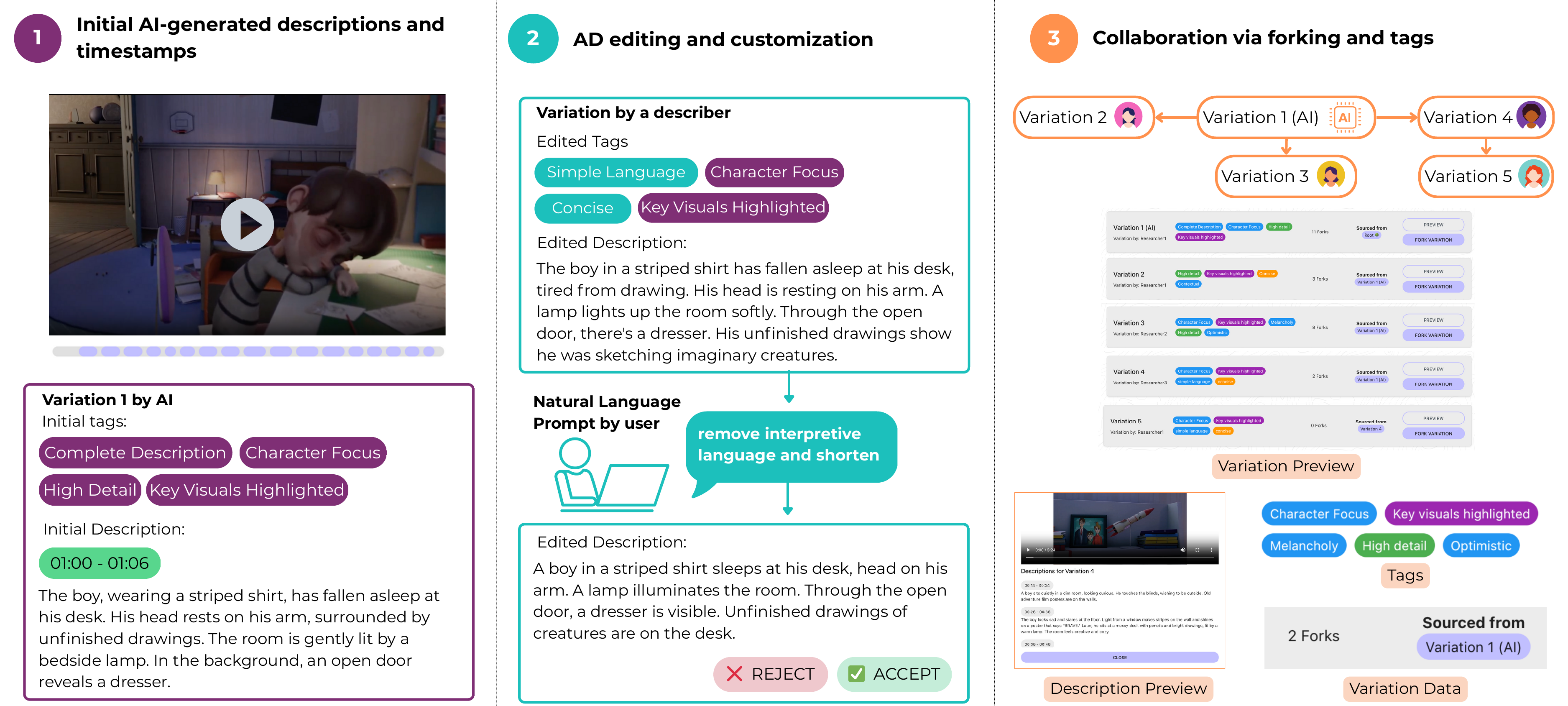}
  \caption{\softwarename\ supports users in creating audio descriptions (AD) through AI generation and prompting and collaboration with other describers. (1) Once a video is uploaded, \softwarename\ creates initial AI descriptions and tags. 
  (2) Users can customize ADs manually or using natural language prompts to edit one or multiple descriptions. (3) Users can collaborate by forking existing AI or human descriptions for a video to create their own variations.} 
  \label{fig:header}
  \Description{DescribePro interface overview illustrating how users create and collaborate on audio descriptions (AD). The figure is divided into three sections: (1) A video is displayed, and a timeline below the video is shown that visualizes where the descriptions are present in the video. DescribePro automatically generates initial descriptions and a set of descriptive tags. A timestamp is shown with one of the generated descriptions. (2) Showcases how users can use a natural language prompt ``remove interpretive language and shorten'' to revise the text. Edited tags and descriptions are also shown. Users can choose to accept or reject the changes. (3) A diagram shows how users can fork an AI-generated variation into new ones (e.g., Variation 2, 3, 4, 5) and label them with different tags (e.g., ``Optimistic'', ``Melancholy''). A preview panel displays the description text for a video.}
\end{teaserfigure}


\maketitle

\section{Introduction}

With the rapid growth of online video platforms, videos have become a popular medium for both entertainment content (e.g., short films) and instructional purposes (e.g., cooking, home repairs). 
This broad usage makes their accessibility important for blind and low-vision (BLV) users through Audio Descriptions (ADs). ADs provide narration of important visual elements in a video, such as characters, actions, facial expressions, and scene changes to support comprehension and engagement~\cite{Walczak2017, Pinnelli2015, Dang2024, packer2015overview}. However, creating high-quality ADs is time-consuming for professionals and challenging for novices~\cite{yuksel2020human}. As a result, ADs are mostly available for high-budget productions~\cite{fresno2022research}, 
leaving much of the online video space, especially user-generated and low-budget content, inaccessible to BLV users \cite{jiang2024}.
Recent studies show that multimodal large language models (MLLMs), when guided by AD guidelines, can generate detailed and accurate visual descriptions~\cite{li2025videoa11y}. However, these models still fail to capture the deeper context and stylistic variation needed for different genres and audiences. This limitation highlights the need for collaborative approaches that combine the efficiency of AI-generated descriptions with the creativity, storytelling ability, and sensitivity to context, tone, and style that human describers provide~\cite{thompson2018audio}. Human describers can add emotion, rhythm, and artistic touches that make ADs more engaging and better suited to each video. This human-AI collaboration maintains both accessibility and high-quality ADs \cite{brashear2021automated, gao2024audio}.

To this end, we present \softwarename, a web-based platform that integrates AI and human collaboration to improve description quality, support scalability, and streamline the AD creation process. 
In \softwarename, each video can have multiple description variations, authored by either AI or humans, each tagged to reflect different styles or purposes. Describers can create a new AD variation by providing custom instructions, which are combined with general AD guidelines, to generate initial descriptions.
Through an interactive interface, describers can prompt the system to revise one or more descriptions iteratively, for instance, by applying global edits across selected scenes to speed up the process. Describers can also fork existing variations to explore alternative narrative styles or collaborate with others. 
By combining human-AI and human-human co-creation, \softwarename\ enables describers to produce accurate, contextually rich, and personalized audio descriptions.

To evaluate \softwarename, we conducted a user study centered on three research questions:

\begin{itemize}
    \item \textbf{\textit{RQ1:}} How do the AI and collaborative features of \softwarename\ support AD authoring? 

    \item \textbf{\textit{RQ2:}} What aspects of existing (AI-generated, human) descriptions do professional and novice describers seek to improve, and how do they use the system to make those improvements?

    \item \textbf{\textit{RQ3:}} How do professional and novice describers differ in using \softwarename?
\end{itemize}

The user study involved 18 participants, including 9 novice and 9 professional describers. After introducing how to use \softwarename, we asked the describers to create and edit audio descriptions for three short videos using the system. The experimental tasks focused on editing AI-generated or human descriptions. Upon completion of the tasks, they filled out a survey on the usability and usefulness of the main interface features. In addition, we measured various metrics during their interaction with the system to assess usage patterns and the system’s impact on description quality and task completion.

Our findings demonstrate that \softwarename\ supports both novice and professional describers in streamlining the audio description workflow through a combination of AI-generated content, interactive editing tools, and community collaboration. Participants rated the system as usable, with an average SUS score of 72.6, and highlighted specific features such as forking, AI-generated baselines, and the prompting interface as useful to varying degrees. Forking was especially valued by both groups, while novices found the AI descriptions helpful in overcoming the initial challenge of starting from scratch. Qualitatively, participants shared that the system reduced the tediousness of AD creation and provided a helpful starting point for writing. 
Most participants expressed optimism about using AI as a collaborative partner in AD authorship, highlighting the potential of \softwarename\ to complement human creativity rather than replace it.
The ability to fork or view other people's variations was seen as a valuable feature. Professional describers recognized its potential to enable customized AD for BLV users, moving from the traditional one-size-fits-all AD approach. Novices, on the other hand, viewed variations as a useful training tool, offering inspiration and guidance as they began their journey in AD creation. In contrast to prior work that largely focused on novice describers, our results offers in-depth insights into how both professionals and novices use AI for AD authoring, highlighting similarities and differences in their needs and perceptions of AI-based AD authoring support. 
We discuss the implications of our results for future work and outline opportunities for leveraging AI and human expertise to support user agency and skill learning. 

In summary, our contributions are as follows:

\begin{enumerate}
    \item \softwarename, a web-based system that facilitates both human-AI and human-human collaboration for creating diverse audio descriptions.
    \item Empirical data that capture how professional and novice describers use and experience AI and collaborative features.
\end{enumerate}

\section{Related Work}

We review prior work on existing approaches and practices for AD and collaboration when authoring AD.

\subsection{Audio Description Practices}
To make video content accessible, the Web Content Accessibility Guidelines (WCAG) suggest adding audio descriptions during silent pauses of a video to describe actions, characters, scene changes, and on-screen text ~\cite{wcag_audio_desc}. Several guidelines have been created for authoring descriptions ~\cite{ofcom2024services, 3play2024AD, dcmp2024descriptionkey, netflix2024guide, ami_dv}. Traditionally, AD authoring has been a manual task requiring describers to craft AD while adhering to guidelines and industry standards. However, the rise of new types of video content, such as short videos (e.g., TikToks, Shorts, Reels), makes it challenging to apply guidelines originally designed for long-form media like films and TV shows.

Research in video accessibility has led to the development of several tools that assist describers~\cite{livedescribe2012, youdescribewebsite, pavel2020rescribe, chang2022omniscribe}. LiveDescribe supported novice describers by identifying silent gaps in the video timeline for inserting descriptions~\cite{livedescribe2012}, while Rescribe helped optimize AD placement by detecting silent moments and automatically adjusting the timing of the descriptions~\cite{pavel2020rescribe}. YouDescribe was built to offer a web-based platform for users to describe YouTube videos~\cite{youdescribewebsite}. Natalie et al.\ introduced a web-based tool that provides automated feedback to help users create high-quality descriptions\cite{natalie2023supporting}. Omniscribe enabled users to produce spatial ADs and immersive labels for 360$^\circ$ videos~\cite{chang2022omniscribe}.
We extend this body of work by introducing collaborative features and interactive AI tools to support editing and refining ADs.

Past research has introduced several tools to support the automatic creation of audio descriptions~\cite{wang2021toward, yuksel2020human, ihorn2021narrationbot, van2024making, ning2024spica}. Yuksel et al.\ developed a system that generated initial AD drafts for sighted individuals to revise into high-quality descriptions~\cite{yuksel2020human, automated_AD}. This was one of the few early studies to explore how AI-generated baselines could support human describers, although it predated the emergence of MLLMs.
More recent work has examined how LLMs can be used to generate automatic textual descriptions. Shortscribe, for instance, produced three levels of hierarchical video summaries for short-form videos~\cite{van2024making}. SPICA enabled BLV users to interactively explore objects in a video and receive AI-generated natural language descriptions~\cite{ning2024spica}. VideoA11y combined MLLMs with AD guidelines to generate descriptions that outperformed novice human annotations and matched trained human annotations in clarity, accuracy, objectivity, descriptiveness, and user satisfaction~\cite{li2025videoa11y}. WorldScribe delivered real-time visual descriptions of live environments, offering customizable options for BLV users~\cite{chang2024worldscribe}.
While these systems focus on fully automated AD creation, AI-generated descriptions are still prone to hallucinations and often fail to capture the right context~\cite{chaoyucvpr}. Recent work has also moved towards seeing AD beyond accessibility and more as an artistic expression~\cite{thompson2018audio, schaeffer2023beyond}, including subjectivity and nuance that is difficult to automate. Additionally, syncing the descriptions with video timing remains a major challenge, reinforcing the importance of a human-in-the-loop approach to AD creation. 

Recently, a few commercial AI-based tools have emerged for AD authoring~\cite{lucariello2024dcmp, audiblesight2024}. Audible Sight generates initial AI descriptions that describers can manually edit~\cite{audiblesight2024}. The DCMP's AI Scene Description Tool generates scene-level descriptions and answers questions to support describers and BLV users~\cite{lucariello2024dcmp}.
However, these tools do not support AI-assisted editing or refinement of the generated descriptions. While they reflect the growing real-world interest in using AI for AD workflows, we still know little about how effective AI-generated text is in supporting human describers during the creation process. 
We extend these efforts to enable AI prompting and co-creation in \softwarename\ and investigate how describers work with AI-generated descriptions and what aspects they seek to improve in descriptions.

\subsection{Collaboration and Customization in AD Authoring}
Creating audio descriptions is inherently a difficult and subjective task as different describers may prioritize different visual elements~\cite{kleege2015audio, fresno2014less}. This can be particularly challenging for novice describers since deciding what to describe can be mentally demanding~\cite{livedescribe2012, fresno2016should}. Previous research has explored how collaborating with BLV users can improve the quality of descriptions~\cite{natalie2020viscene, jiang2023beyond, jiang2022co}. For example, Viscene, a web-based platform, allowed novice describers to get feedback from blind and sighted reviewers on scene descriptions~\cite{natalie2020viscene}. Jiang et al.\ highlighted the collaborative approach for 360$^\circ$ videos, showing how BLV AD creators and experts contributed unique insights that might be missed by sighted describers~\cite{jiang2023beyond}. Similarly, Pationaki advocated for close collaboration between describers and artistic teams when creating AD for dance and musical performances, to enhance the audience's experience with more immersive and accurate representation~\cite{patiniotaki2022audio}. 
These studies highlight the value of collaboration in the AD creation process. However, to our knowledge, no work has examined how describers themselves can collaborate during the creation of descriptions.
In this work, we investigate how the collaborative features of \softwarename\ support the AD authoring experience of describers.

Prior work suggests that BLV users have different preferences for AD consumption. These preferences are often shaped by their level of visual acuity ~\cite{chmiel2022homogenous, cheema2024describe}, and the type of video content they are consuming~\cite{jiang2024, natalie2024audio}. Recent research highlights the limitations of the traditional one-size-fits-all approach to AD and emphasizes the need for offering different levels of detail and supporting customization~\cite{natalie2024audio, Walczak2017}. 
While platforms like YouDescribe allow multiple volunteers to describe a video, they lack collaborative editing features. Describers cannot modify existing descriptions and must start from scratch. 
We build on the need for AD variation and collaboration by gathering qualitative feedback from describers on how \softwarename's \emph{variation} and \emph{tagging} features can support the creation of multiple, customized descriptions for BLV users.

\begin{figure*}[htbp] 
    \centering
    \includegraphics[width=1\textwidth]{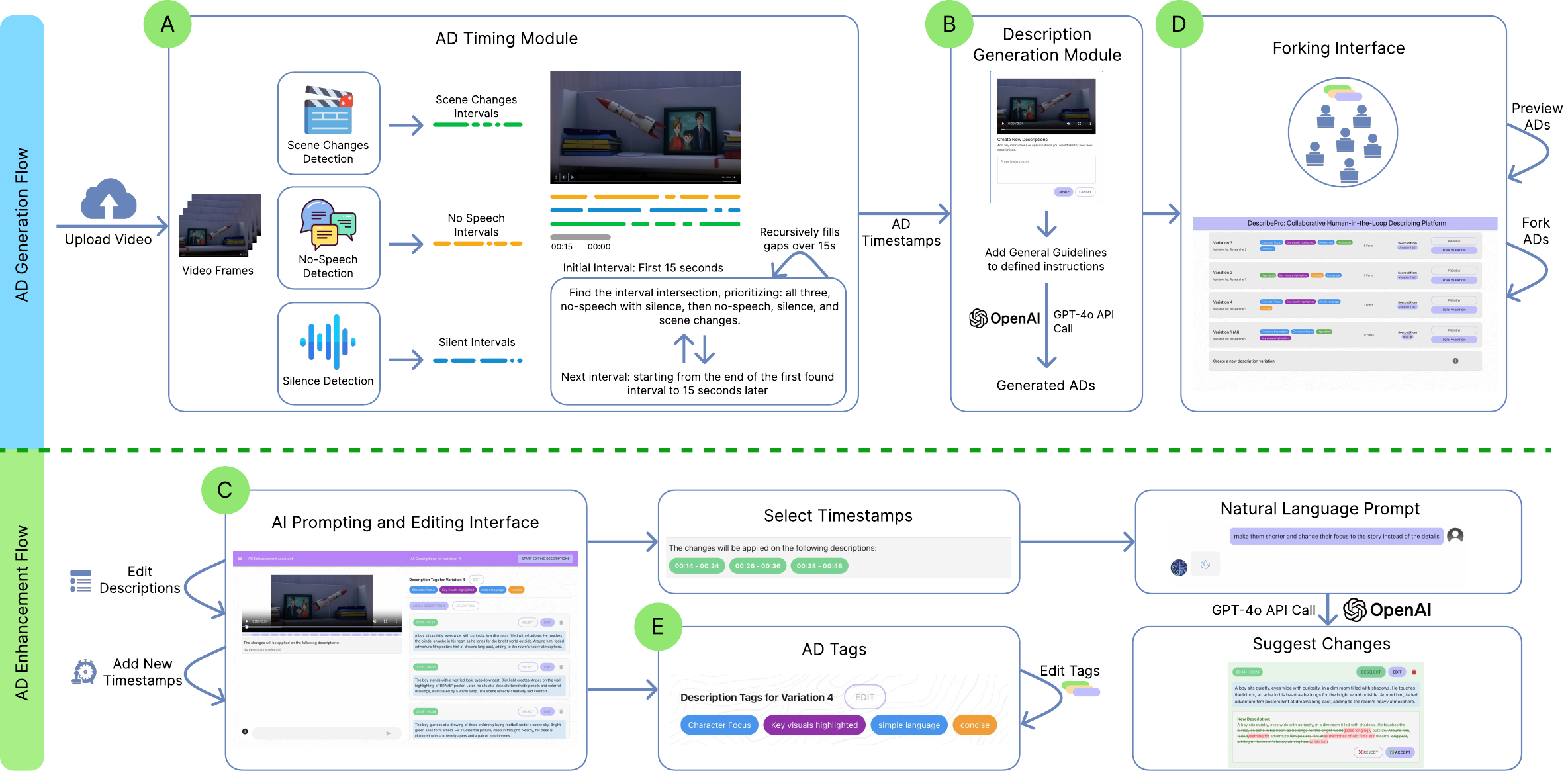}
    \caption{
    Overview of the \softwarename\ workflow, comprising the AD Generation Flow (top) and AD Enhancement Flow (bottom).
    In the AD Generation Flow, users begin by uploading a video to the \textit{DescribePro} platform. (A) AD Timing module automatically identifies silent segments, speech, and scene changes, prioritizing intervals that intersect silence and no-speech periods. These are used to generate structured audio description (AD) timestamps. Next, (B) Description Generation Module processes the video frames and produces initial ADs using GPT-4o.
    (C) AI Prompting and Editing Interface allows describers to review, select timestamps, and edit descriptions through natural language prompts or manual input. Describers can preview the revised descriptions from GPT-4o and iterate to produce high-quality, accessible ADs. (D) The Forking interface enables previewing and creating alternatives from existing variations. Finally, (E) AD Tags can be viewed and edited through the interface. 
    }
    \label{fig:describepro-flow}
    \Description{A figure showing the DescribePro system architecture, split into two sections: the AD Generation Flow (top) and the AD Enhancement Flow (bottom). The Generation Flow includes modules for detecting silence, no-speech, and scene changes to determine suitable AD intervals. These are visualized as color-coded bars along a video timeline. The Description Generation Module connects to GPT-4o to create initial descriptions. In the Enhancement Flow, users interact with an editing interface that allows timestamp selection for AI prompting via natural language and manual text editing. The open AI edits the descriptions for the selected timestamp. AD tags are shown, which are editable. Each section is annotated with letters (A–E) to indicate key workflow steps.}
\end{figure*}

\section{DescribePro}

\softwarename\ is a web-based platform for human-human and human-AI collaboration on AD authoring (Figure \ref{fig:describepro-flow}). 
Users can upload videos to generate initial AI descriptions, refine them through AI prompting or direct edits, and collaborate with others on the platform. To support these capabilities, \softwarename\ includes the following components: (A) AD Timing module, (B) Description Generation module, (C) AI Prompting and Editing Interface for revising descriptions, (D) Forking Interface to create and compare multiple AD variations, and (E) AD Tags, which label content and stylistic differences across versions. The system uses GPT-4o~\cite{hurst2024gpt} as its core AI model.

\subsection{Generating Descriptions for a Video}
When the user uploads a video, \softwarename\ first runs the \emph{AD Timing} module to identify key timestamps for inserting descriptions, then uses the \emph{Description Generation} module to generate the initial AI descriptions for the corresponding video frames.

\subsubsection{\textbf{AD Timing Module}} 
This module processes video content to determine ideal timestamps for inserting audio descriptions, following established AD guidelines. It extracts the audio track and video frames, then analyzes them to detect three key signals: (1) silence, (2) no-speech segments, and (3) scene changes.
Silence refers to the complete absence of sound, whereas no-speech specifically indicates the absence of human speech, even if other sounds like music or ambient noise are present. This distinction helps avoid interrupting important dialogue (primary focus) while enabling the insertion of ADs or pauses during no-speech moments, such as instrumental breaks in music videos, though this is not always feasible in content with continuous audio.
Further, scene change detection analyzes visual frames to mark meaningful transitions and ensure that descriptions align with significant visual shifts.
\softwarename\ uses lightweight, open-source libraries for these tasks: Pydub~\cite{robert2018pydub} is used for audio signal analysis and silence detection, the Silero Voice activity detection (VAD) model~\cite{Silero} for identifying speech versus non-speech segments, and PySceneDetect~\cite{Castellano_PySceneDetect} for detecting scene changes in the video.

As shown in Figure~\ref{fig:describepro-flow}, the AD Timing module analyzes the video in 15-second segments to identify intervals where silence, non-speech, and scene changes overlap. These overlapping regions indicate ideal moments for inserting audio descriptions—i.e., natural pauses in audio that coincide with meaningful visual transitions.
If a full overlap of all three signals is not found, the module selects partial overlaps, prioritizing combinations of all three intervals, intersection of silence and no-speech, followed by no-speech, silent gaps, and scene changes. Finally, if any of the final timestamps exceed 15 seconds in length, they will be recursively split at their midpoints to allow for ADs insertion, as they are sufficiently long. This hierarchical approach ensures that selected timestamps are well-suited for audio descriptions, balancing both auditory and visual cues.

\begin{figure*}[htbp] 
    \centering
    \includegraphics[width=\textwidth]{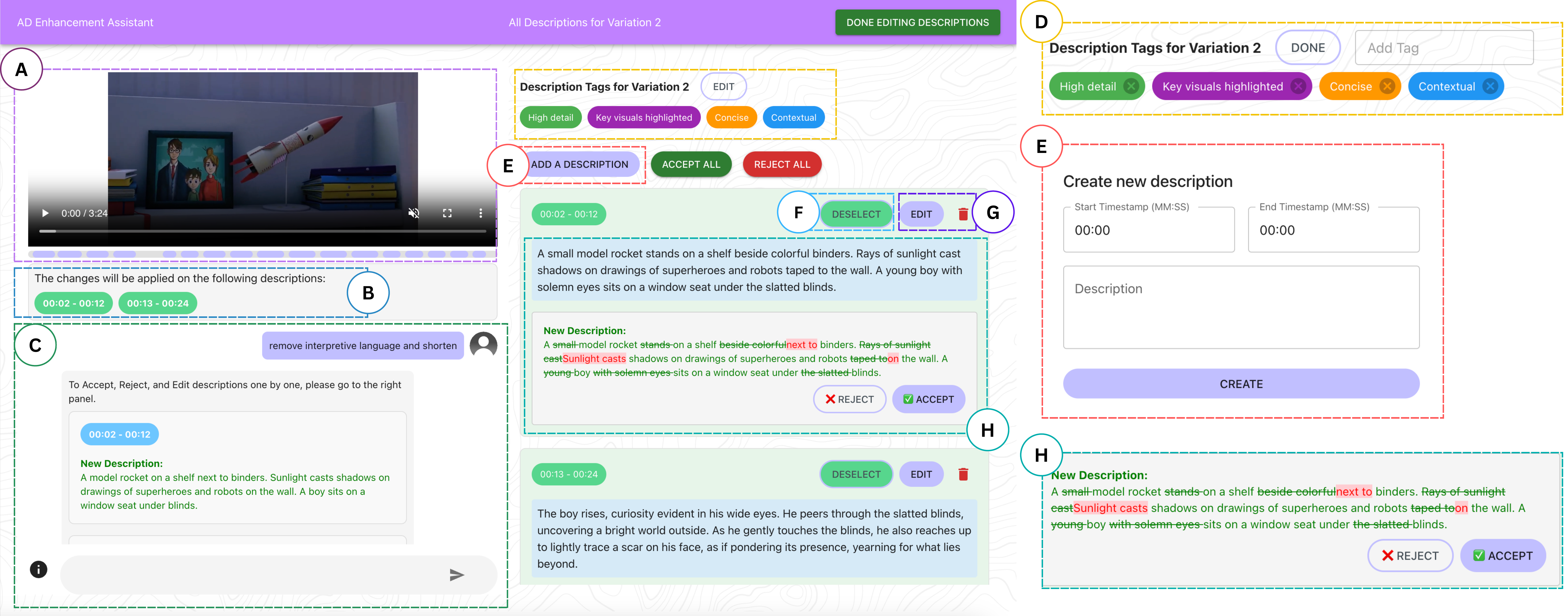}
    \caption{The AI Prompting and Edit interface supports human–AI collaborative editing of video descriptions. The left side shows the main interface components, while the right side provides a zoomed-in view of selected elements. (A) The video player and timeline display descriptions aligned with visual moments for easy previewing. (B) Key timestamps selected for prompt-based editing. (C) Prompt box where users can instruct the AI to revise audio descriptions. (D) Editable tags that describe the AD, aiding in organization and filtering. (E) ADD A DESCRIPTION button for creating new scene descriptions. (F) SELECT/DESELECT button to indicate whether a description is selected for editing. (G) EDIT button to manually modify or delete a description. (H) Description box displaying both the original and AI-generated versions based on user prompts. Edits in the new description are highlighted, and users can accept or reject the changes.}
    \label{fig:prompting-interface}
    \Description{Interface mock-up illustrating DescribePro’s editing view with labeled zones: a video preview player, selected timestamps display, a prompt box for natural language prompting to AI for editing the descriptions, editable AD tags, and old vs. new AI descriptions side-by-side for visually tracking changes. Includes buttons for accepting, rejecting, and editing descriptions, supporting a user-guided editing workflow.}
\end{figure*}

\subsubsection{\textbf{Description Generation Module}}
Because writing ADs from scratch can be difficult, \softwarename\ assists describers by generating initial descriptions as a starting point. After uploading a video, the describer is provided with an input box to enter custom instructions, which guide the AI on what to emphasize, how to structure the descriptions, and which elements to prioritize. The system then combines these custom instructions with a curated set of 42 general AD guidelines from AD training resources. This approach builds on prior work showing that GPT prompted with these 42 AD guidelines produces descriptions that outperform novice describers and are comparable to trained annotators in clarity, accuracy, objectivity, descriptiveness, and user satisfaction~\cite{li2025videoa11y}. To ensure the output aligns with the describer’s intent, the prompt is designed to prioritize user-provided instructions over the general guidelines (see Appendix~\ref{appendix:video-prompt} for the full prompt).

Next, the module generates audio descriptions by sending a prompt and a sequence of video frames to the GPT-4o model.
To prepare the input, the module samples one frame every two seconds and constructs a sequence of frames between consecutive key timestamps. These frame sequences are then batched and sent to GPT-4o, along with the prompt, to produce the corresponding descriptions. For subsequent batches, previously generated descriptions are included alongside the prompt and video frames to provide better context and minimize repetition.
Finally, the module saves the generated descriptions, along with their associated timestamps and frame sequences, to \softwarename’s database.

\subsection{Editing Descriptions for Quality and Style Preferences}
Describers can view existing descriptions for a video using \softwarename's \emph{AI Prompting and Edit} interface and revise them through AI prompting and direct edits (Figure \ref{fig:prompting-interface}).

\begin{figure*}[htbp] 
    \centering
    \includegraphics[width=\textwidth]{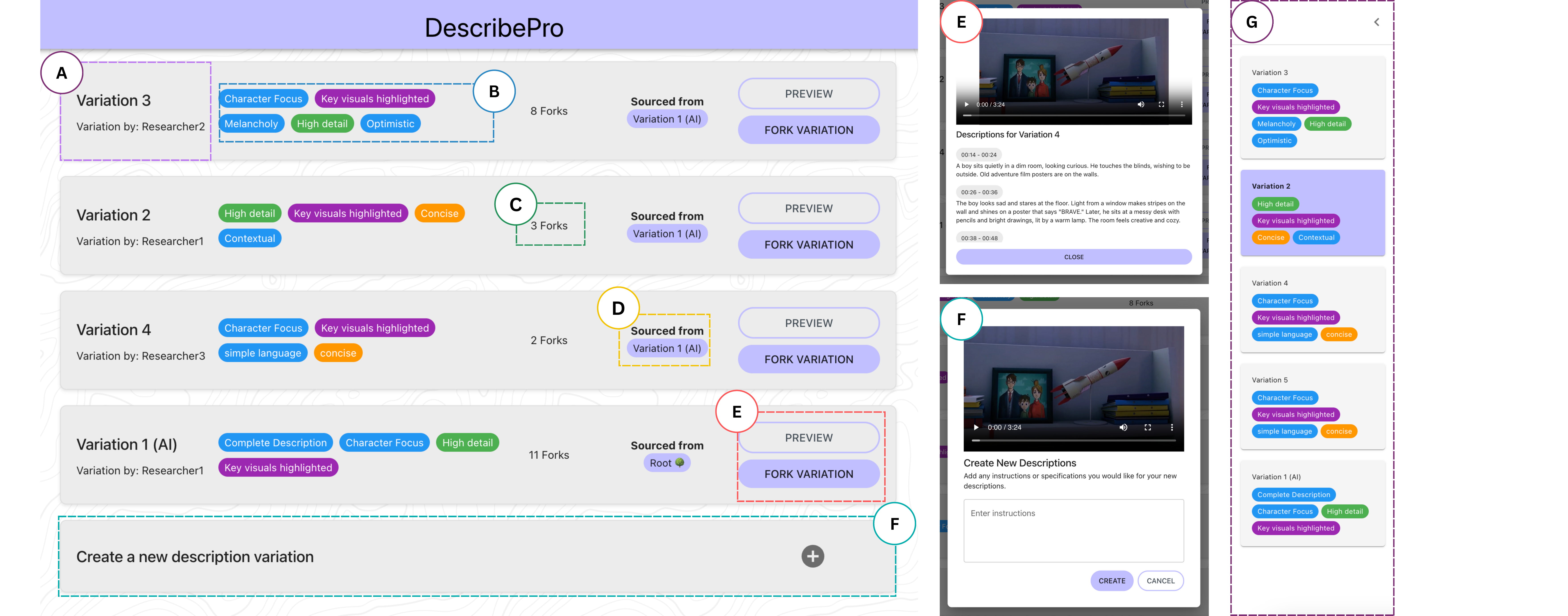}
    \caption{The Forking interface displays the initial AI-generated description (Variation 1) and subsequent human-created variations, supporting collaborative refinement. (A) Variation header showing the name of the variation and its contributor for clear attribution and edit tracking. (B) Tags that communicate the narrative purpose and style of each variation at a glance. (C) Fork number indicating how many times a variation has been copied, serving as a measure of reuse and popularity. (D) Source lineage referencing the parent variation to support traceability of iterative changes. (E) PREVIEW and FORK VARIATION buttons that let users view or build upon existing descriptions. Clicking PREVIEW opens a pop-up displaying the video and corresponding descriptions to help users decide whether to fork. (F) Pop-up window for creating a new variation from scratch using AI. (G) Navigation bar on the AI Prompting and Edit page for accessing and previewing existing variations while editing descriptions.}
    \label{fig:variatons-page}
    \Description{Variations page screenshot showing the forking workflow in DescribePro. Includes a list of variations with tags, fork counts, and source attribution. Preview and fork buttons allow users to view descriptions for that variation and branch from existing variations to create an editable copy, respectively. The preview modal, the modal for creating new AI variations, and a vertical navigation bar for quick access to each variation are displayed.}
\end{figure*}

\subsubsection{\textbf{AI-Powered Prompting}}
A key feature of \softwarename\ is its AI-powered prompting interface, which enables describers to revise multiple ADs at once using an MLLM. Describers can select one or more key timestamps and apply targeted or bulk modifications across several descriptions simultaneously.
For example, they can prompt the system to add visual detail, adjust the level of interpretation (e.g., describing a character’s emotion), shorten the text, or include/exclude specific elements like people’s appearances or background objects. This flexible customization supports rapid refinement of ADs to suit different stylistic preferences and accessibility needs, enabling the creation of tailored variations for diverse audiences.

On the backend, the user’s prompt, along with the selected timestamps and corresponding video frames, is sent to the MLLM to generate revised descriptions (see Appendix~\ref{appendix:input-prompt-prompt}). 
Once the revised descriptions are generated, the prompt is stored, and the updated descriptions are returned in the response. In the interface, each revised description is shown directly below the original, aligned with the corresponding key timestamp.
To help users compare versions, differences such as insertions and deletions are visually highlighted using color and text formatting (Figure~\ref{fig:prompting-interface}). Users can accept or reject revisions individually or in bulk, and the system stores all accepted revisions while logging the number of approvals and rejections for tracking and analysis.

\subsubsection{\textbf{Direct Editing and Iterative Refinement}}
\softwarename\ also supports manual editing, allowing describers to refine audio descriptions without relying on AI. Moreover, describers can adjust timestamps, add or delete descriptions, and make detailed edits to ensure better synchronization with the video. This full manual control enables precise customization, similar to traditional AD authoring tools like YouDescribe, while maintaining flexibility for diverse stylistic or accessibility needs. All manual edits are saved to the database and automatically logged by the system.

\subsection{\textbf{Collaboration among Human Describers}}
With \softwarename, describers can build on each other's work by \emph{forking} existing descriptions to create new variations. The system displays a customizable set of \emph{tags} for each variation to highlight its content and stylistic features, making it easy to distinguish between different AD versions for the same video (Figure~\ref{fig:variatons-page}).

\begin{table*}
\footnotesize
  \caption{Description of the 18 describers in our study. Participant numbers indicate their experience: Professional (P) and Novice (N).}
  \label{tab:Demographic Information}
  \centering
  \begin{tabular}{p{0.3cm}p{0.3cm}p{1.3cm}p{1.3cm}p{6.3cm}p{3.3cm}}
    \toprule
    \textbf{P\#} & \textbf{Age} & \textbf{Race} & \textbf{Gender} & \textbf{Prior AD experience} & \textbf{Used AI for AD creation} \\
    \midrule
    P1 & 35 & White         & Non-Binary & 6+ years of experience in describing live performances & No\\
    P2 & 37 & White         & Female     & 8 years of experience in describing film and TV & No\\
    P3 & 66 & White         & Female     & 24 years of experience in describing live theater, museum exhibits, art installations, and more  & No\\
    P4 & 26 & White         & Non-Binary & 4+ years of experience describing film, TV, educational videos, commercials, and opera & No\\
    P5 & 52 & White         & Male       & 10 years of experience performing and producing AD & Yes - ChatGPT, DCMP application\\
    P6 & 29 & Asian, White  & Non-Binary & 5 years of providing live and prerecorded AD for live performances  & No\\
    P7 & 45 & White         & Male       & 10+ years of post-production AD experience for a media outlet for BLV individuals & No\\
    P8 & 42 & White         & Non-Binary & 20 years of experience recording, mixing, producing and writing description & No\\
    P9 & 67 & White         & Female & 25 years of experience in AD for live theatre, films,  short videos, TV, artwork, and live events & No\\
    \midrule
    N1 & 22 & Black, White  & Female     & Edited and improved AI descriptions for a project  & Yes - ChatGPT\\
    N2 & 55 & White         & Male       & Took classes in audio description but have yet to work professionally & No\\
    N3 & 56 & White         & Female     & Read a book on AD and interested in learning more & No\\
    N4 & 45 & White         & Female     & Took an overview class on AD & No\\
    N5 & 47 & White         & Female     & Took brief training in audio description & No\\
    N6 & 54 & White         & Female     & Took training workshops, but currently not working in AD creation & No\\
    N7 & 35 & White         & Female     & Research as an academic, using audio description in/for dance performances & No\\
    N8 & 27 & Asian         & Male     &  Research on audio description for various video genres & Yes - GPT-4o\\
    N9 & 61 & White         & Female     & Research on the linguistic features of audio descriptions & No\\

    \bottomrule
  \end{tabular}%
  \label{tab:participants}
\end{table*}

\subsubsection{\textbf{Forking to Create Variations of Existing Descriptions}}

A key collaborative feature of \softwarename\ is its forking mechanism, which allows describers to create alternative AD variations by building on existing work, whether originally authored by others or generated by AI.
Describers can browse existing ADs for a video, preview their content, and review associated metadata such as tags and fork count. The number of forks can serve as a signal of quality or popularity, helping users choose strong starting points and avoid redundant variations.
When a describer selects a variation to fork, \softwarename\ increments the fork count, copies relevant data (e.g., descriptions, timestamps, tags), assigns a new variation name, and links the variation to the user.
The new forked variation then appears in the \emph{Forking} interface, ready for further editing and customization.

\subsubsection{\textbf{Customizable Tags for Variations}}
To help distinguish between AD variations for a video,
\softwarename\ features a customizable tagging system that highlights the focus and style of each version. These tags make it easier for both describers and BLV users to quickly find descriptions that best meet their needs. For example, a tag might indicate whether a description is concise or detailed, focused on the main storyline, or emphasizes visual elements such as characters’ appearances or the environment. Tags can reflect any aspect the describer considers important, offering a quick snapshot of what each variation prioritizes.

\softwarename\ automatically assigns initial tags to each AD variation, but describers can revise or expand these tags as needed.
To create structured and meaningful tags, we compiled a predefined list of keywords (see Appendix~\ref{appendix:tags-prompt}), informed by prior research on audio descriptions~\cite{natalie2024audio,cheema2024describe,schaeffer2023beyond} and our own work with BLV users.
When a variation is created, the system sends its descriptions to GPT-4o, which selects the four most relevant tags from the predefined list and may suggest up to two additional custom tags.
The predefined tags ensure consistency across variations, while the custom ones allow flexibility to capture unique aspects.
Once generated, tags are stored in the database and shown on both the \emph{Forking} and \emph{AI Prompting and Editing} interfaces.
This guarantees that each variation is meaningfully categorized, making it easier to browse and compare.
Describers retain full control and can modify the tags at any time to better reflect the characteristics of their AD.


\section{User Study}
To understand how \softwarename{} supports describers in their AD authoring workflow, we conducted a user study with 18 participants, including 9 novice and 9 professional describers.


\subsection{Paticipants}
We recruited 9 novice and 9 professional describers for the user study through online groups (e.g., Reddit, Facebook) and snowball sampling (Table~\ref{tab:participants}). Professional and novice describers were compensated \$50 and \$30 respectively, to reflect their levels of experience. 
Participants with prior paid experience in creating descriptions and providing feedback were categorized as professional describers. To verify their background, we asked professional describers to provide a sample of their AD work before the session. In contrast, novice describers were familiar with AD guidelines and best practices (e.g., through training workshops or research) but had no prior paid experience in creating descriptions.
The participants ranged in age from 22 to 67. Professional describers had over 12 years of experience creating audio descriptions, on average.
Only a few participants reported using AI in their AD creation workflow ($n=3$).

\begin{figure*}[htbp] 
    \centering
    \includegraphics[width=0.9\textwidth]{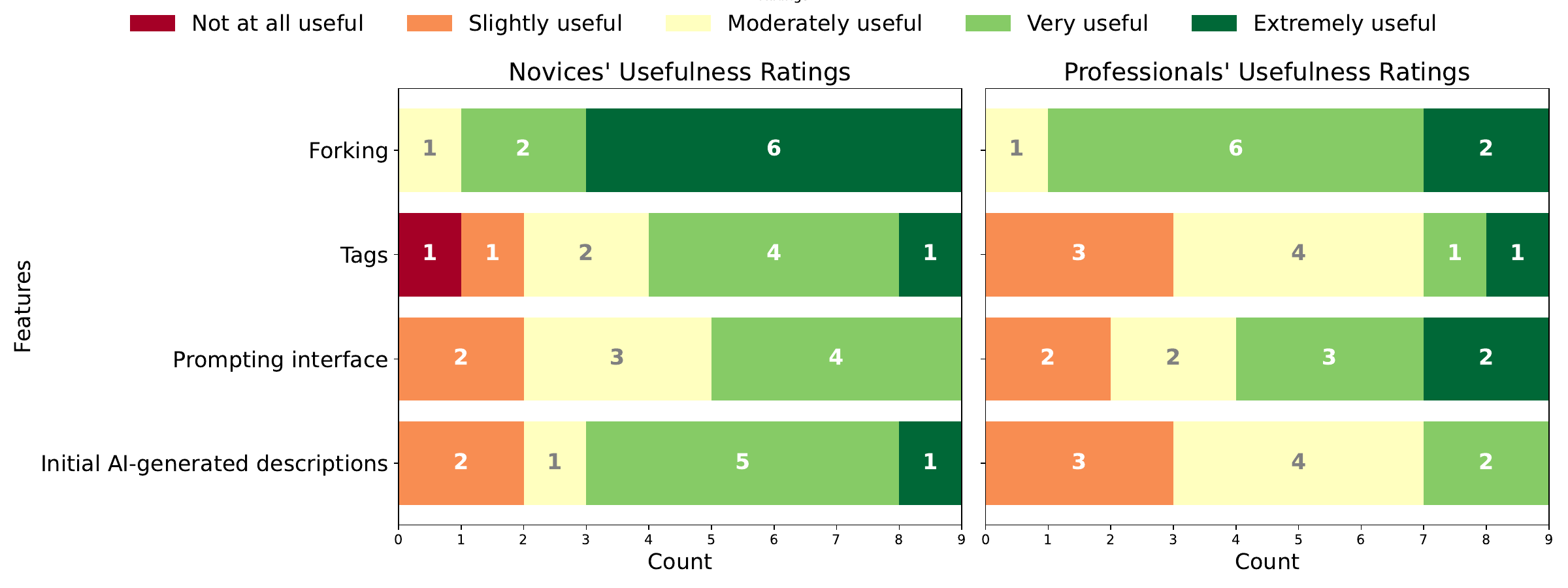}
    \caption{Distribution of usefulness ratings for \softwarename\ features by novice (left) and professional (right) describers.}
    \label{fig:ratings}
    \Description{A pair of horizontal stacked bar charts titled ``Novices' Usefulness Ratings'' and ``Professionals' Usefulness Ratings'' show the distribution of usefulness ratings for four DescribePro features: Forking, Tags, Prompting Interface, and Initial AI-generated Descriptions. Each feature is rated on a 5-point scale ranging from ``Not at all useful'' to ``Extremely useful,'' represented by different colors. Both novices and professionals rated ``Forking'' highly useful, with most selecting ``Very'' or ``Extremely useful.'' Novices gave lower ratings to ``Tags,'' while professionals gave moderate ratings across most features. Each bar is labeled with the count of responses per rating level.}
\end{figure*}

\subsection{Procedure}
The study session was organized into four phases: (1) an introduction phase that included a demographic survey and an overview of the study, (2) a main task phase with three AD creation tasks using \softwarename, (3) a post-questionnaire phase, and (4) a semi-structured interview.
After completing the demographic survey, participants received a brief orientation to \softwarename\ and its features. They then performed three AD tasks designed to evaluate the interface's support for human-AI interaction and collaboration.
In the first task, participants refined a set of AI-generated descriptions.
In the second task, they reviewed three existing AD variations, each created by a member of the research team, and selected one to revise.
The third task was open-ended, allowing participants to either build on a researcher-created AD variation or start from scratch with AI-generated descriptions.
To reduce learning effects and account for differences between videos, we counterbalanced the order of the first two tasks and the video assignments across participants.
Following the tasks, participants completed a post-questionnaire that included the System Usability Scale (SUS)~\cite{brooke1996sus} and ratings on the usefulness of various interface features.
Finally, participants took part in a semi-structured interview to share detailed feedback on their experience using \softwarename for AD creation. All interviews were audio-recorded and transcribed for analysis.

\subsection{Videos} 
We selected six short videos (each 3–4 minutes long) for the three tasks in our user study. For the first two tasks, we selected one entertainment and one instructional video based on the top requested genres by BLV users on YouDescribe~\cite{youdescribewebsite}: a short film\footnote{\url{https://www.youtube.com/watch?v=iD_tsK_aqIQ}} and an instructional cooking video~\footnote{\url{https://www.youtube.com/watch?v=e7pLKhI7l08}}. For each video, we provided three human-edited AD variations, created by members of our research team by revising baseline AI-generated descriptions in accordance with standard AD guidelines.
For Task 3, participants could choose from four different videos: a short film clip\footnote{\url{https://www.youtube.com/watch?v=0Taa1-Exd8I}}, a how-to origami tutorial\footnote{\url{https://www.youtube.com/watch?v=cZdO2e8K29o}}, a dog show segment\footnote{\url{https://www.youtube.com/watch?v=F-0iepogz1g}}, and a scenic tour of Santorini\footnote{\url{https://www.youtube.com/watch?v=MFOWMbrp5FM}}. Each of these videos included two variations, one human-edited and one AI-generated. Participants were free to select which video to work on and which variation to fork and revise.

\subsection{Data Analysis}
For the quantitative data, we recorded the time participants took to select a variation and the time they spent editing each video. We also tracked the number of prompt-based edits participants performed, along with how many of those edits were accepted. Moreover, we analyzed the content of the prompts entered by participants. One author developed an initial closed-code codebook to capture variation in prompt content (see Appendix~\ref{appendix:codebook}). Then, two authors independently applied the codebook to the full set of prompts. After the initial round, they achieved an inter-coder agreement of 73.1\% exact matches and 17.9\% partial matches (i.e., cases where two codes were applied and one overlapped). The remaining 9\% of the prompts had differing codes. All discrepancies were discussed and resolved collaboratively to finalize the coding. We prompted GPT to list unused guidelines; all 42 guidelines appeared in at least 65\% of descriptions. Unused guidelines were contextually irrelevant (race or character details in an instructional video). 
For the qualitative interview data, we used thematic analysis following Braun and Clarke’s approach~\cite{clarke2021thematic}. Two authors independently applied open coding to five interview transcripts using MAXQDA qualitative analysis software. After coding the initial interviews, they met to compare codes and discuss overlaps and differences. One author then coded the remaining transcripts. This author drafted an initial set of themes, which both authors reviewed and refined through discussion, using the coded data for reference. This process resulted in four final themes.

\section{Results}
We present the quantitative results first, followed by the qualitative themes derived from the interviews.

\subsection{Quantitative Results}

\subsubsection {\textbf{Usability and Usefulness Ratings}}
The average SUS score was 72.6 ($SD = 13.9$), slightly above the benchmark of 68 for web-based interfaces~\cite{bangor2008empirical}, suggesting positive perceptions of \softwarename.
Figure~\ref{fig:ratings} presents the 
ratings from novice and professional describers on the usefulness of \softwarename\ features. Novices rated forking and the initial AI-generated descriptions as more useful than professionals, probably because 
editing existing content is often more approachable than starting from scratch. Professionals found greater value in the prompting interface, possibly due to their experience with crafting and refining AD content. Forking was the most useful feature overall, while tags received the lowest ratings and the greatest variation in perceived usefulness. 


\subsubsection {\textbf{User Prompts and Edits}}
\begin{figure}[htbp]
    \centering
    \subfigure[Frequency of codes for types of AI prompts used by novice and professional participants.]{
        \includegraphics[width=0.48\textwidth]{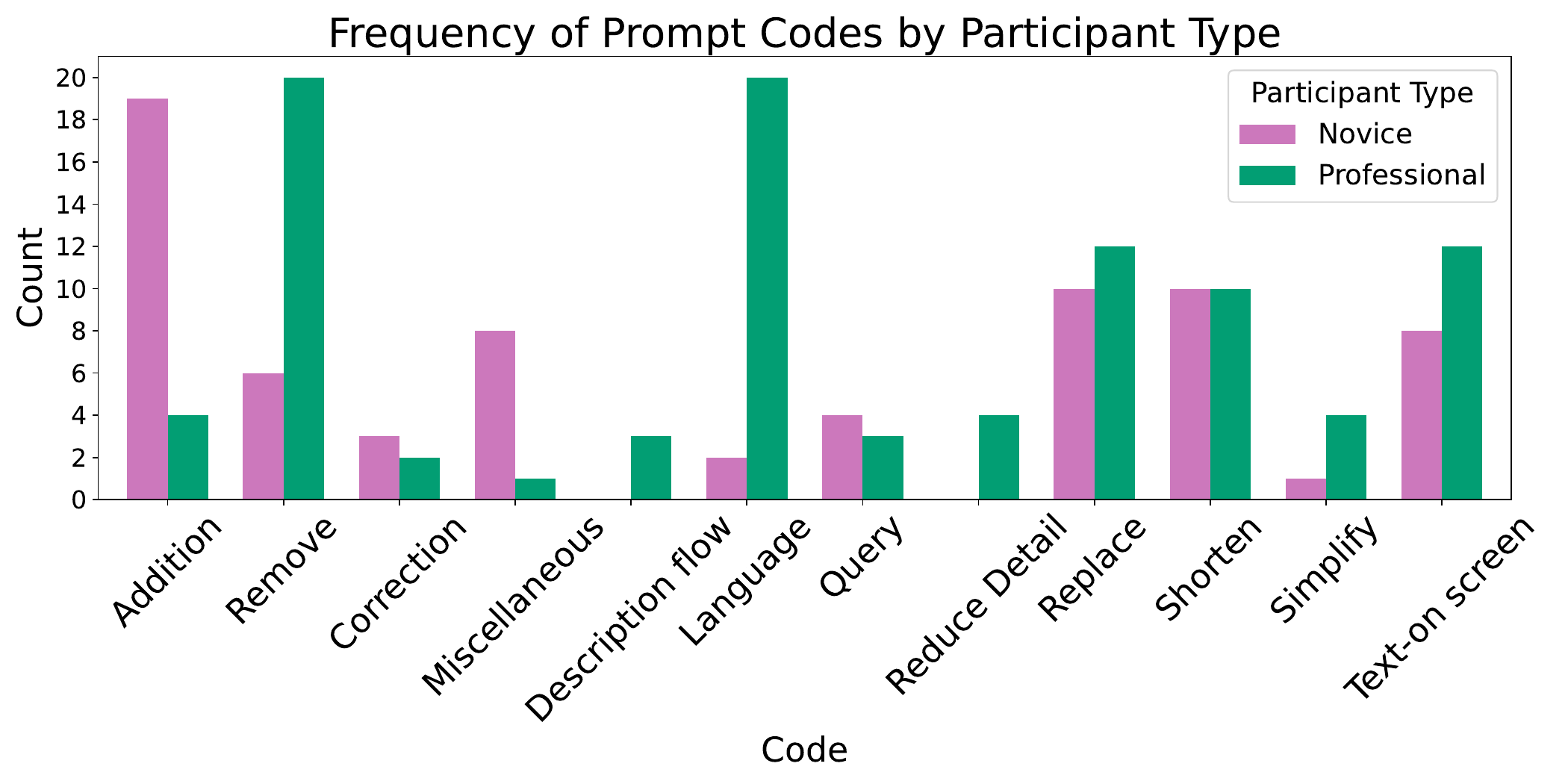}
        \label{fig:edit_counts}
    }
    \hspace{0.1cm}
    \subfigure[Frequency of codes for types of AI prompts based on task type]{
       \includegraphics[width=0.48\textwidth]{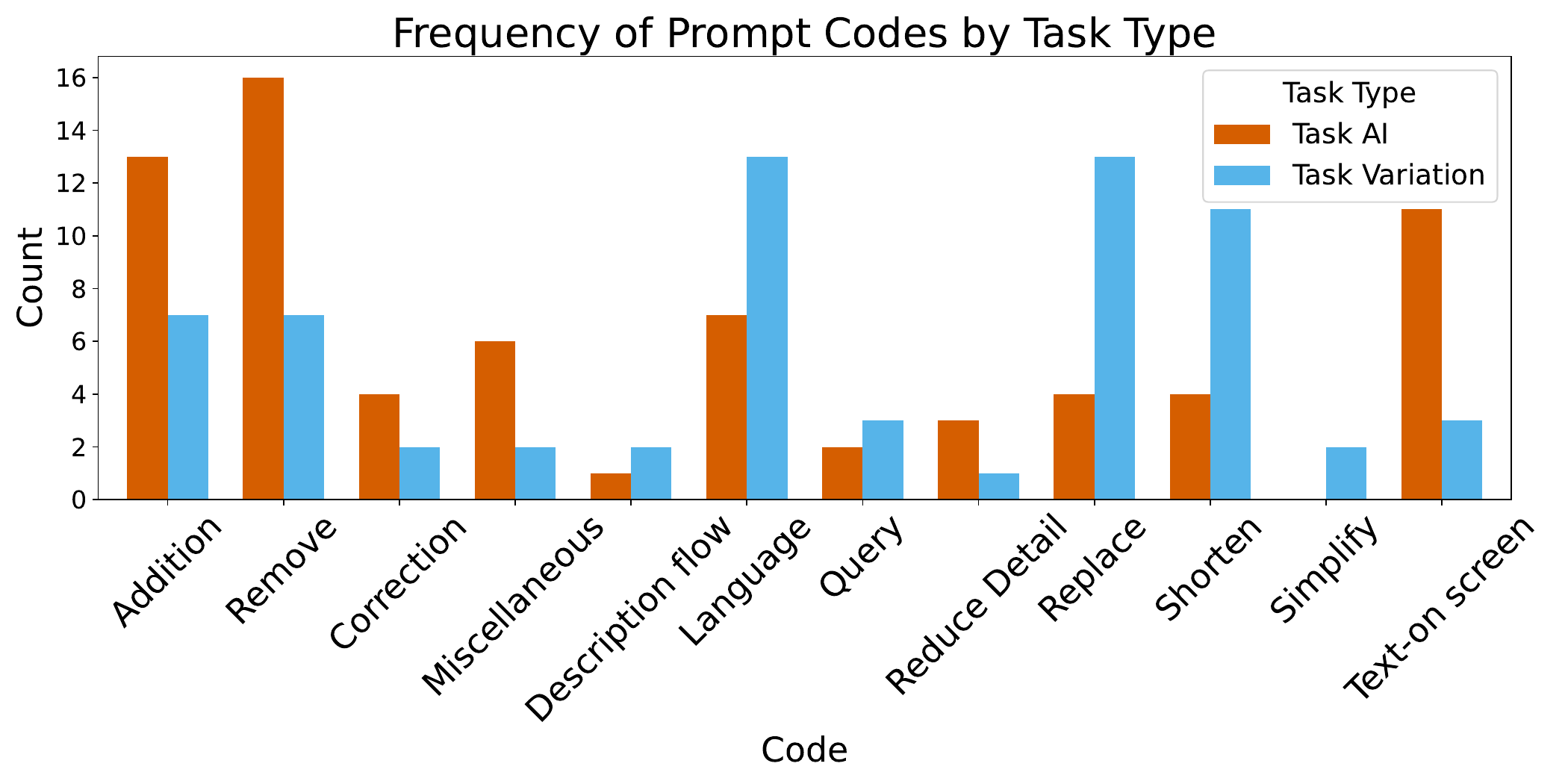}
        \label{fig:time_taken}
    }
    \caption{Frequency of prompt codes categorized by (a) participant expertise and (b) task variation source. (see~\ref{appendix:codebook} for code definitions.)}
     \label{fig:codes}
\Description{Two bar charts showing the frequency of AI prompt codes used in DescribePro. The top chart compares prompt types between novice and professional participants. Novices used more addition and miscellaneous prompts, while professionals used more removal and language-focused prompts. The bottom chart compares prompt types between AI and human variations. Participants used more addition prompts for AI variation, and more shortening and refinement prompts for human variation, indicating differing editing strategies.}
\end{figure}

Figure~\ref{fig:codes} shows the frequency of prompt codes categorized by participant type (novice vs.\ professional) and task variation (AI-generated vs.\ human). The data reveals distinct editing styles between the two groups. Professionals more frequently used prompts that focused on language refinement (e.g.,\emph{``make passive to active verbs''}) or the removal of unnecessary details (e.g.,\emph{``remove mention of the kitchen''}).  In contrast, novice describers were more likely to use prompts to add descriptive content (e.g. \emph{``Add description after `greek flags waving in the distance.'''}). 
Differences also emerged based on the type of variation participants worked on. Prompts related to adding or removing content were more common when editing AI-generated descriptions, as were prompts for including on-screen text. In contrast, they used ``shorten'' prompts when editing human-generated content. Language refinement prompts were also more common for human variations. Across all tasks, professional describers used the AI prompting feature 82 times, compared to 65 times for novice participants. Professionals also demonstrated a higher acceptance rate of the AI’s revised descriptions (72.8\%) compared to novices (60.1\%). This suggests that professionals were more effective in translating their editing intentions into successful AI prompts—likely due to their greater experience and ability to craft precise instructions that aligned with their descriptive goals. Finally, participants adjusted 11.4\% of timings, with an average shift of about 0.8s, suggesting initial timings were generally accurate.

\subsubsection{\textbf{Revised AD Length and Edit Time for AI vs. Human Variations}}Participants tended to shorten descriptions more when working on the human variation. Figure~\ref{fig:edit_counts} illustrates the change in total word count by task type. We conducted a mixed ANOVA to examine the effects of task type (AI vs.\ Human) and participant type (Novice vs.\ Professional) on changes in description length. The main effect of task type was significant: $F(1, 16) = 0.729$, $p = .006$, $\eta_p^2 = .380$, indicating that participants made greater modifications to the human-generated descriptions compared to the AI-generated ones. In addition, the Levenshtein distance, measuring the number of single-character edits needed to transform one string into another, was computed for the original and modified descriptions. On average, descriptions modified for human variations involved more edits than those in the AI variation, with a higher average Levenshtein distance (Human Variation: 136.7, AI variation: 51.4), reflecting more insertions (Human Variation: 22.2, AI variation: 11.6), deletions (Human Variation: 84.5, AI variation: 26.9), and substitutions (Human Variation: 30.0, AI variation: 12.9) per description.
 One possible explanation is that the human variations were more verbose, as they were created by members of the research team who were not professional describers. In particular, novices often include too many details, even when attempting to follow AD guidelines~\cite{fresno2016should}.

Figure~\ref{fig:time_taken} shows the total time taken to complete each task. Overall, participants spent a similar amount of time on the AI variation task ($M = 23{:}44$, $SD = 09{:}23$) and the human variation task ($M = 21{:}12$, $SD = 08{:}24$).
These results suggest that AI-generated descriptions can serve as a strong starting point for editing. Supporting this, an equal number of participants chose to work from AI-generated and human variations in Task 3, indicating that many found the AI output to be a viable foundation for further refinement.

\begin{figure}[htbp]
    \centering
    \subfigure[Reduction in word count when editing description]{
    \includegraphics[width=0.48\textwidth]{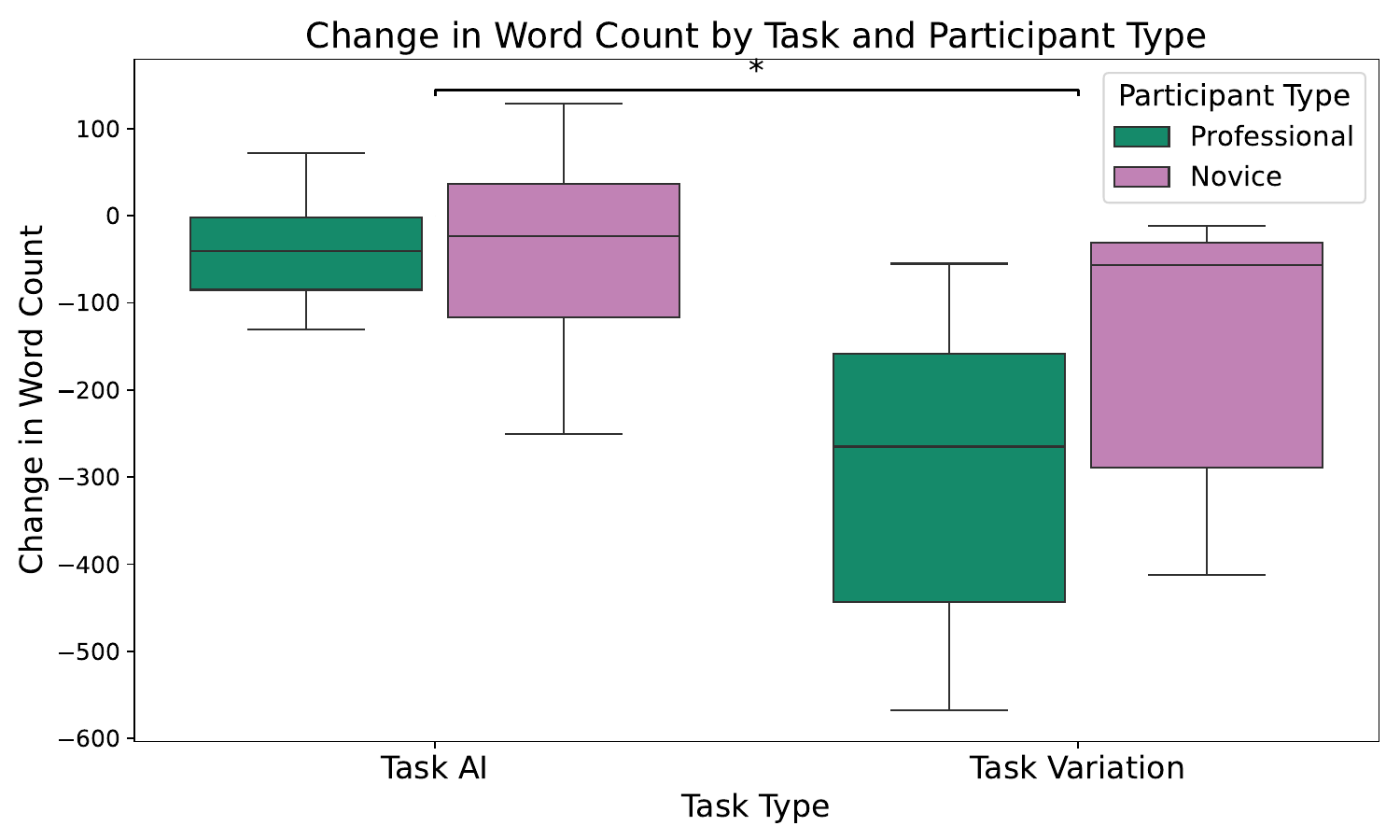}
        \label{fig:edit_counts}
    }
    \hspace{0.1cm} 
    \subfigure[Time taken by participants to complete editing descriptions]{
    \includegraphics[width=0.48\textwidth]{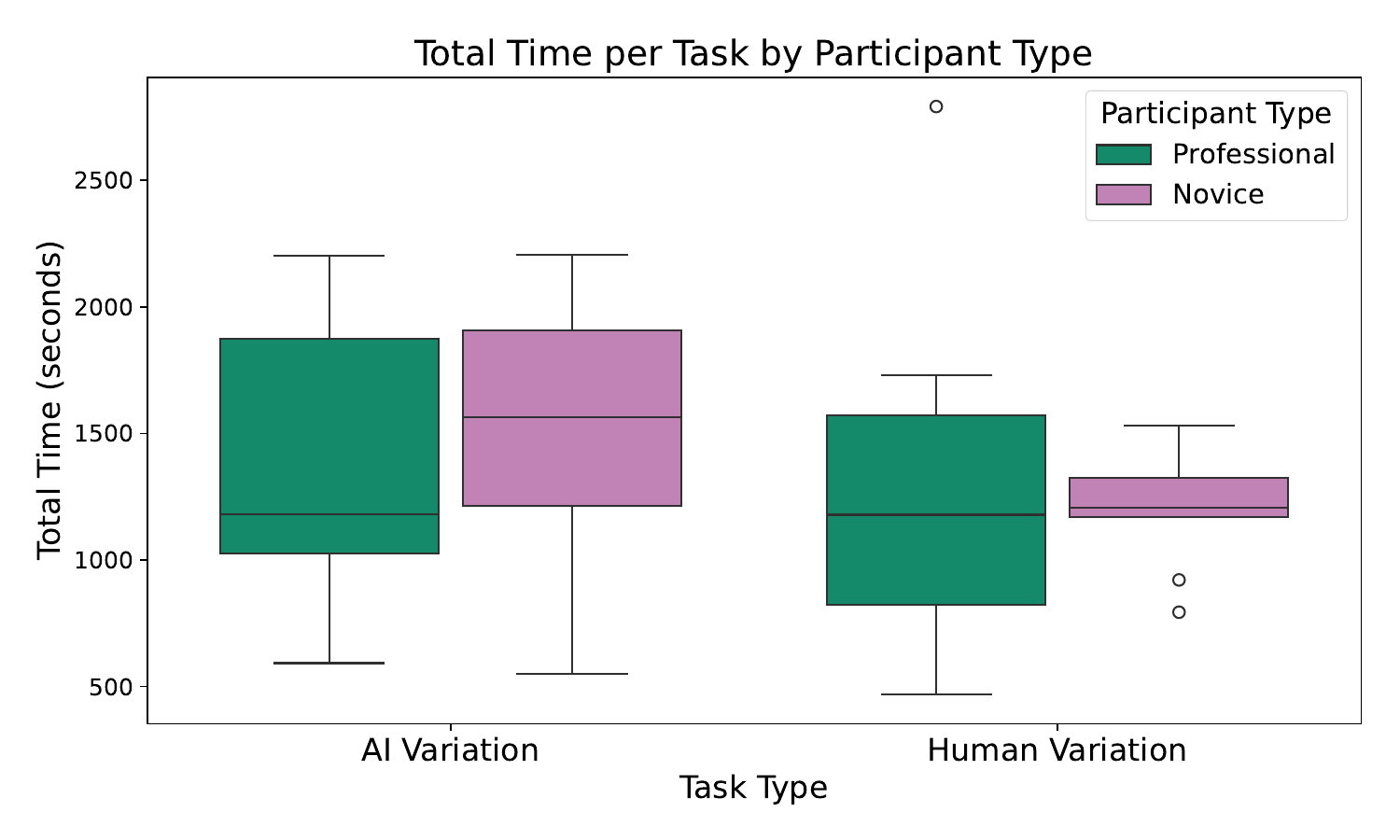}
        \label{fig:time_taken}
    }
    \caption{Figure~\ref{fig:edit_counts} shows the reduced word count for participants for their edited descriptions. The negative y-axis implies the edited descriptions were shorter than the original descriptions. Figure~\ref{fig:time_taken} shows the total time taken by participants to complete the task. The total time includes the time it took to choose a variation and the time it took to edit it.}
    \label{fig:editing}
\Description{Two box plots showing differences in editing behavior for AI and human variation editing tasks. The top plot shows the change in word count by task type. The plot indicates that both novice and professional participants reduced the word count more when editing human variations. Professional describers, in particular, made greater reductions in word count compared to novice describers. The bottom plot shows the total time it took for participants to edit the AI variation or the human variation. AI variation tasks generally took more time to complete, especially for novice participants. Professional describers showed greater variability in time taken for both tasks compared to novices.}
\end{figure}

\begin{table*}[h]
\centering
\caption{Average similarity scores between original and revised ADs across semantic, lexical, and stylistic metrics for novice describers and professional describers.}
\label{tab:similarity-metrics}
\begin{tabular}{llcc}
\toprule
\textbf{Metric Type} & \textbf{Model / Method} & \textbf{Novice Describers} & \textbf{Professional Describers} \\
\midrule
Lexical    & SequenceMatcher (\texttt{difflib}) & 0.650 & 0.580 \\
Semantic   & Cosine Similarity of BGE-M3 Embeddings & 0.882 & 0.837 \\
Stylistic  & Cosine Similarity of LUAR-MUD Embeddings & 0.866 & 0.826 \\
\bottomrule
\end{tabular}
\end{table*}

\subsubsection{\textbf{Comparing Original and Revised ADs across Similarity Metrics}}

To evaluate the extent of modifications made by novice and professional describers, we used metrics that capture the lexical, semantic, and stylistic differences between the original and revised descriptions (Table~\ref{tab:similarity-metrics}). For lexical differences, we applied a word-level similarity measure using Python’s SequenceMatcher module from the difflib library, which compares the description text by matching subsequences and computes the ratio of matched to total sequence lengths. For semantic similarity, we used the BGE-M3 model~\cite{li2023bge}, which generates embeddings that encode sentences into 1024-dimensional vectors. To assess how closely the semantic content of the descriptions aligned, the cosine similarity between embeddings for original and revised ADs was calculated. For stylistic similarity, we applied the LUAR-MUD model~\cite{rivera2024luarmud}, which analyzes sentence structure and phrasing patterns and converts text into 512-dimensional vector representations, quantifying style alignment across sets of sentences. Similarly to the semantic analysis, cosine similarity was applied to these vectors to evaluate the degree of style alignment between the original and revised ADs. According to all three metrics, professional describers’ revisions were consistently less similar to the original descriptions than those of novices (by 4–7\% depending on the metric). This suggests that professionals tended to make more substantial edits. However, the relatively high similarity scores for both groups indicate that the original descriptions still served as a solid foundation for authoring ADs.


\subsection{Qualitative Results}
\begin{figure*}
    \centering
    \includegraphics[width=\textwidth]{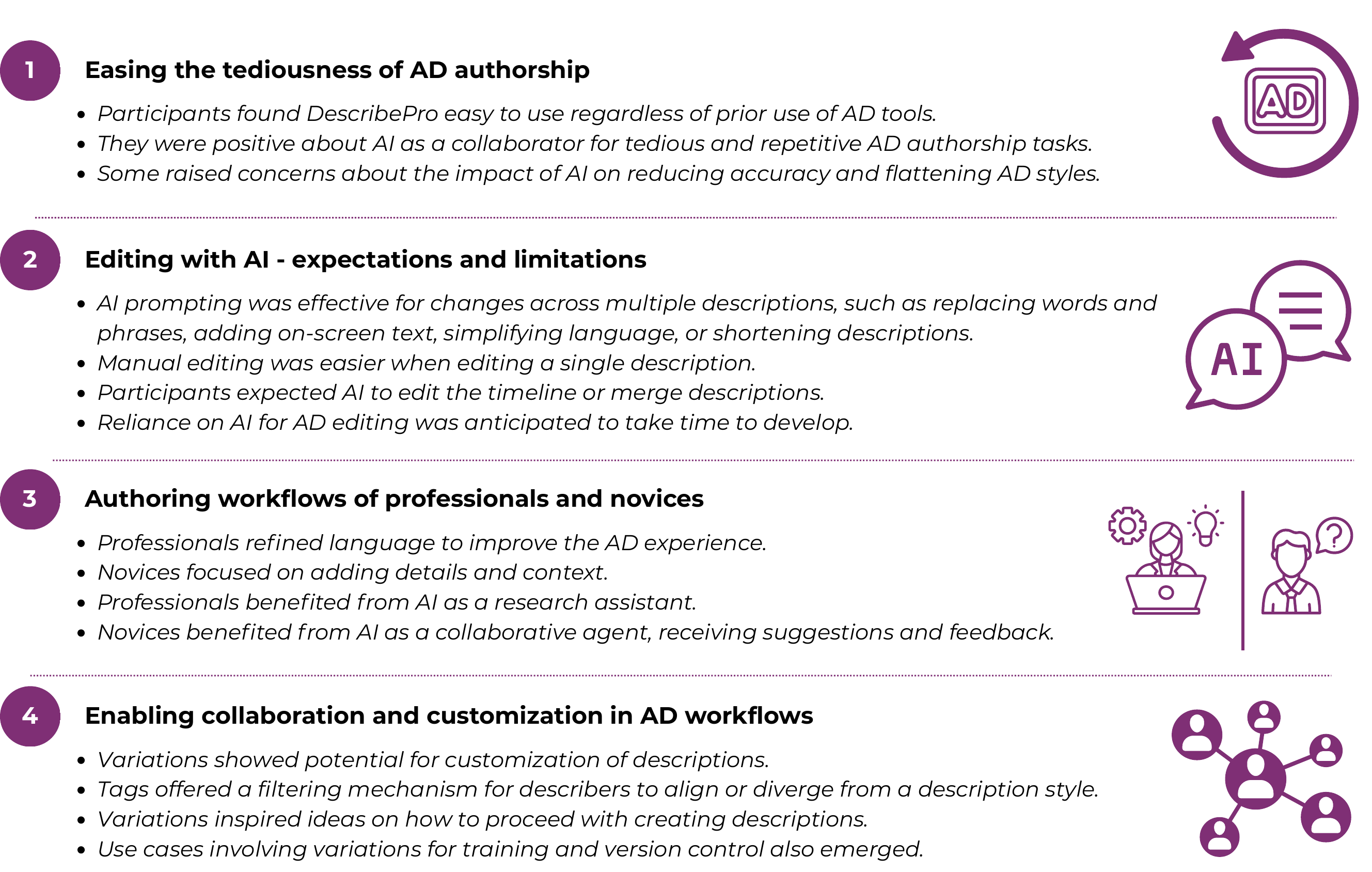}
    \caption{Overview of the four themes and sub-themes from interviews with professional and novice describers about their experience with \softwarename.}
    \label{fig:themes}
    \Description{Overview of four major themes and sub-themes derived from interviews with novice and professional audio describers using DescribePro. The figure is structured into four numbered sections, where the title represents the theme, bullet points represent the sub-themes, and a purple icon illustrates its concept (e.g., document for authorship, AI speech bubble, human figures for collaboration)}
\end{figure*}

We identified four themes based on the participant interviews and feedback (Figure~\ref{fig:themes}).

\subsubsection{\textbf{Easing the Tediousness of AD Authorship}}
Participants expressed how \softwarename{} reduced the physical and cognitive effort of creating descriptions, especially in the early stage of drafting descriptions, highlighting the potential of AI-assisted tools to streamline the description workflow for novice and professional users.
Most participants highlighted that \softwarename\ was easy to use and quick to learn ($n=11$). The overall reaction was positive,  with several finding it ``interesting'' and a novel approach to tackling challenges of AD creation. As N4 remarked: \emph{``I think it's an interesting concept to solve a problem of, you know, audio description taking too long. You know, there is a lot of content [to describe] out there.''}. 

Most professional participants felt ``incredibly positive'' towards \softwarename\, expressing that AI-generated descriptions provide a good starting point for describers ($n=9$). Several mentioned how AI can help reduce the tedious and repetitive tasks of creating descriptions, such as transcribing on-screen text and visual details. As P5 highlighted: \emph{``Getting those baseline visual descriptions is so helpful. But then, from there, that's where the human element has to come in... this application takes away a lot of the repetitive uselessness, like there's so much tediousness in [creating] audio description that's not helpful''}. However, a couple of professional describers felt differently, noting that based on their current workflow (P4), or when describing content they are familiar with (P3), it is easier to start from scratch than to edit existing descriptions: \emph{``I think that having a kind of full generated script, it's not helpful for me, just with my process.'' (P4)}.

Several novices also liked \softwarename, highlighting how the descriptions reduced some of the mental demands associated with AD authorship. The AI descriptions alleviated the burden of starting from scratch ($n=6$), as N5 highlights: \emph{``I thought it was helpful because it gave me a starting place, you know, from which to edit, instead of having to come up with something from scratch.''}. In addition to easing the initial writing process, AI-generated descriptions also helped highlight visual details that participants might otherwise have overlooked, as N4 expressed: \emph{``I found out the basketball was part of the story ... I watched the video, but I didn't notice that detail the first time I watched it''}. Beyond offering content support, \softwarename\ also serves as a confidence-building tool for newcomers in AD creation. As N3 reflected: \emph{``it gave me confidence in where to start with for the audio description''}.

Despite the perceived benefits, professional describers raised important concerns about the accuracy and subjectivity of AI-generated content ($n=2$). Describers might often forget to remove hallucinations from AI descriptions: \emph{``If AI generates something inaccurate, and then the editor misses that, that could be a potential issue... it's just kind of hard to keep track of what's there when it didn't come from your own mind.} (P8). P7 expressed how AI might not yet be able to capture some of the nuances around describing race or identifiable sounds, but still acknowledged the flexibility of not being \emph{``locked into one sort of definitive description.''} In addition, P5 expressed that AI might remove some of the subjectivity from AD and potentially ``flatten'' the descriptions. In contrast, some describers noted that they could edit the AI descriptions to match their style. For example, P2 mentioned that after prompting AI to remove emotional language and making direct edits \emph{``[the description] felt a bit more like my voice.''}  
Similarly, several professionals were optimistic about the use of AI for AD authorship ($n=5$). This is largely because they felt confident that human discernment cannot be replaced by AI, as P1 highlights: \emph{``I think the way forward isn't necessarily letting AI do all the description. It's a collaboration between the human describers and the AI to help generate it. So we don't have to write everything from scratch every single time.''} 


\subsubsection{\textbf{Editing with AI: Expectations and Limitations}}
Participants found AI prompting useful for widespread changes ($n=6$), such as replacing certain words, removing interpretive or emotional language, shortening descriptions, or adding on-screen text. P7 mentioned how this is a more accessible way of interacting with AI, additionally mentioning how it can be used to customize descriptions to focus on the environment or character based on video viewing intent. Even when the describers did not accept AI's output, they often used part of the edited descriptions. As N3 described \emph{``The on-screen text part was too verbose in my opinion, so I asked it to make it smaller. But then it went and edited the whole thing instead of just the text, and I didn’t like that, so I edited it myself. But I used the shorter wording for the on-screen text from the chat in my edit. So that was helpful to have the results, even if I didn’t directly use them.''} This suggests that AI prompting can act as a support tool for describers in ideation and customization, offering alternatives or information that describers can use to refine descriptions.

For editing a single description or making more precise changes, participants preferred to manually edit the descriptions. These adjustments were often more difficult to convey through prompting, as P4 summarizes the challenge: \emph{``My brain doesn't really know what to do with like translating my changes to an AI conversation. It felt like more work and more time to try to tell it what to do, and then see if it was right, and then accept it or reject it.''} Often, participants had to alter their prompt for the desired outcome. The AI also occasionally made edits to parts of the description that were not targeted. In such cases, participants also resorted to manually editing descriptions. To mitigate this, some participants proposed solutions such as adding a ``refine'' button for the AI-revised description so the user can create multiple alternatives of the edited descriptions (P1) or allowing word or phrase level accept or reject within each description to give them more control over AI edits (P5). 

In addition, in a few cases, there was a mismatch between the editing capabilities using the chat feature and the edits a user wanted. For example, N6 attempted to delete selected descriptions by prompting the AI, rather than using the delete button. Similarly, N8 tried to merge selected descriptions into one description through a prompt. These examples suggest new forms of edits that the interface should support through AI prompts. On the other hand, it could also reflect a learning curve. A couple of participants acknowledged that relying on AI for edits would take time, particularly because most had not used AI in their AD workflows before (see Table~\ref{tab:Demographic Information}). Despite the challenges with prompt formulation, participants at all experience levels saw value in AI as a supportive tool, particularly to generate alternatives and aid in ideation.

\subsubsection {\textbf{Authoring Workflows of Professionals vs. Novices}}
Based on qualitative feedback, there was variation in workflows between professional and novice audio describers. When asked about the changes made to the descriptions, all participants unanimously identified adding on-screen text, adding context, and removing redundancies from baseline AI descriptions. However, there were differences in the way participant groups approached editing ADs with AI assistance. 
Professional describers focused on refining language and would benefit from an AI agent that supports research. They often approached the tasks with a clear understanding of what edits they would like to make in the descriptions. In addition to the general edits made by both groups, professional describers also mentioned improving the ``experience'' (P4), the language style, and the general description flow ($n = 3$). This was also evident in their AI prompting, often testing linguistic edits, such as removing excessive adverbs, replacing passive voice, and removing emotional language to create more neutral descriptions. Professional describers also focused more on creating short and concise descriptions. For example, when choosing between a variation to edit, P8 chose one that they considered was ``less wordy''. Beyond textual editing, professional describers emphasized how gathering context for a video often extends beyond what is visible in the video. As P8 explains: \emph{``And you're watching the Westminster dog show. You are gonna want all the details. You are gonna want to know the exact name of the breed. So that would be types of things that I would go and research, I would look up this dog, I would look up its past competitions, and like there's so much you can dig into.''} This suggests that professionals might benefit from an agent serving as a research assistant, especially when dealing with describing content that is less familiar to them. 

Novice describers prioritized detail and sought collaborative AI support. Compared to professionals, they focused more often on adding context and detail to descriptions (professional = 2, novice = 4). This is because novice describers often worry about missing out on important visual elements. Also, novice describers often needed support beyond textual editing. Their prompts included requests such as ``Find better timing'' (N6), or queries like ``can you tell me what's the language on screen that isn't English?'' (N7). Moreover, N2 imagined a more interactive conversational agent that could offer corrective feedback, as explained: \emph{``If it could distinguish that, I mean, if it knew better than I did ... Or if it even asked me as a prompt, `You described this as a black eye. I think it may be a birthmark. What do you think?' [that feedback] kind of gives me the option if it feels I was in error. (N2)''}. This suggests that novice describers may benefit the most from a co-creation agent that not only follows commands, but also helps with decision making by offering suggestions, clarifications, and corrective feedback. Such a system could help foster greater confidence in novices during AD authoring.

\subsubsection{\textbf{Enabling Collaboration and Customization in AD Workflows}}

The initial intention behind having different variations was to explore participants' different narrative styles. However, several other use cases also emerged, including tracking description versions, creating alternative variations, and training novices in description creation.
Forking, the ability to duplicate, edit, and iterate over a copy of a variation, was considered a valuable feature by several participants ($n=8$). The use of forking for AD customization for BLV users was recognized and valued by several participants ($n=5$). Variations were seen as a useful tool, especially for educational videos, as different variations can enhance understanding. N5 highlighted: \emph{``people then can select you know what they would like to view or to view it multiple times to get a different take on it, and maybe enhance their understanding. (N5)''} Similalrly, N7 who participates in agility competitions, highlighted the value of being able to create multiple variations tailored to user familiarity with the video content, as she explains: \emph{``I think the more variation[s] the better... if we go with the agility [dog show] example... I would write extremely specific for someone that has the lingo and the jargon to understand the sport, and then one that could be more of a laid back, more neutral for someone that is not into the competition.''} The use of variations extended beyond traditional video content. In the context of theatrical performances, forking was viewed as a way to curate different tracks, for example, descriptions focused on lighting, movement, or set design. P6 illustrated this creative use case: \emph{`` So I work a lot in performance. And so sometimes I'll make different track for the same video, and like one track will be just light. One track would be just movement, and then one track will be like set design. And so the forking was helpful if I was to take it in that direction to be like, okay, I can get like base information and then get specific, depending on like, what track an audience member might want to pick.''}

Variations supported learning and collaboration by enabling participants to view how others approached a description. This feature was particularly appreciated by novice describers, who found it useful to view and reuse others' work for the same task ($n = 4$). For example, N4 used the navigation bar (Figure~\ref{fig:variatons-page}-G) to preview descriptions created by other users and copied the on-screen text from one of those versions for cooking instructions in their own variation. Novices also emphasized the value of learning from others' edits, as N5 stated: \emph{`` if you're just starting out, it's useful to see what other people wrote to help you learn the ropes and give you ideas.''} This suggests that variations can serve as a training tool ($n=4$). 
Professional describers found variations useful for collaboration in team-based settings. Rather than using it for inspiration, they saw variation as a means of version control and documentation across AD creation stages. As P2 described: \emph{``So let's say that someone on my team is writing, and then I edit and like do like a [quality control] QC Pass, and then someone afterwards an extra QC pass ... it's good to have like records of each one of those''}. 


Tags describing the variations were often overlooked by participants ($n=7$), but showed potential as a filtering selection for both describers and BLV users ($n=6$). A few participants used the tags to choose which description to fork from ($n=3$). For instance, P6 described how a tag shaped their perception of a variation: \emph{``One of the Brave (video) ones said 'optimistic', which, as someone who tries to describe without adding too much like interpretation, I thought that was kind of funny. And I could see how it showed up in the description, and so part of my work ended up like taking out some of the optimism.''} This suggests that tags can serve as criteria to shape the intent of revisions to diverge or further align with a specific description style. Additionally, participants felt that tags can be a useful mechanism for BLV users to filter which variation to choose from: \emph{``I think it opens up a whole new world. Audio description has been so binary—does it have AD or not? What you're doing with tags is saying: not only does it have AD, but you can have it this way, or you can focus on this.'' (P5)} N8 also suggested that it might be better for describers to select from a set of predetermined tags rather than adding open-ended tags. This could make it easier for BLV users to know what to look for, as it would standardize their interpretation. Several participants also noted that tags might be a good source for describing the video content rather than just the AD variation. This would allow BLV users to skim both the video and AD content based on tags. P1 suggested that tags could guide not only variation selection, but also instruct the AI generation of the AD: \emph{``So you can say this one is more artistic—I want more of the detailed, flowery language. Whereas this one is more instructional, so it needs to be more objective-based''}.  These comments highlight the promise of different collaborative features in AD authorship that are often missing from current AD systems.

\section{Discussion}
We present \softwarename\ and investigate the following research questions: (RQ1) How do the chat and collaborative features of \softwarename\ support AD authoring? 
(RQ2) What aspects of existing (AI-generated, human) descriptions do describers seek to improve? (RQ3) How do professional and novice describers differ in using \softwarename? For RQ1, we identified the benefits and challenges of using chat-based prompting to edit descriptions. Participants appreciated chat for editing multiple descriptions efficiently, while manual editing was preferred for precise, single-description changes. The Forking feature received the highest usefulness ratings, and participants valued the flexibility offered by variations and the potential of tags. Additionally, use cases such as within-team collaboration and using variations to train novice describers also emerged from the feedback.
Regarding RQ2, all participants aimed to remove redundancies and enhance contextual information in the descriptions. Overall, describers made more reductions to the human-written variation than the AI-generated one, as reflected in changes to description lengths and their prompting behavior. Several participants noted that the amount of editing work felt similar regardless of whether they began with AI or human descriptions, a finding supported by the comparable editing durations across both types.
For RQ3, professional describers frequently focused on linguistic refinements to improve clarity, tone, and flow, aiming to enhance the AD experience for BLV users. In contrast, novice describers concentrated on adding detail and context, often expressing concern about missing key visual information. Novices found variations helpful for inspiration and learning, seeing them as a potential training tool. Meanwhile, professionals valued variations for version control and offering BLV users a choice in description style. These results uniquely compare the workflows of professional (n=9) and novice (n=9) describers using an AI tool for AD authoring, making this work one of the largest studies involving professional describers in accessibility research.

Based on our findings, we discuss how collaborative and AI-assisted AD tools can support training for novices, enable control and flexibility in co-creating AD, and inform the development of AD guidelines for various video genres. Finally, we reflect on the social and ethical implications of using AI in accessibility workflows and outline the limitations of our study.

\subsection{Supporting Active Learning in AD Tools}
\softwarename{} or similar AI-assisted AD tools can function as interactive training environments for novice describers by offering structured support and reducing cognitive load. Traditional AD training methods, such as workshops or written guidelines, remain underdeveloped~\cite{fryer2016introduction, mazur2021audio}, and novices often struggle to apply these guidelines in practice. For example, they may find it difficult to identify silent gaps or follow AD conventions when creating descriptions. This raises the question: Can AD authoring tools accelerate hands-on learning for beginners?
Prior research shows that description quality improves when novices use templates~\cite{yuksel2020human, morash2015guiding} or receive feedback on their work~\cite{natalie2020viscene, natalie2021efficacy}. Our findings further support this by showing that initial descriptions help reduce the cognitive effort of starting from a blank slate.
Moreover, AD is inherently subjective as describers often prioritize different visual elements~\cite{kleege2015audio}. In this context, the ability to view and compare multiple description variations becomes a valuable learning opportunity. Seeing how others approach the same task can help novices develop a personal style, understand tradeoffs in description choices, and engage in peer learning as part of the authoring process.

Furthermore, automatic feedback has been shown to enhance the descriptiveness of ADs created by novices~\cite{natalie2023supporting}. However, novice describers often need support across multiple dimensions, including the accuracy of the description, adherence to AD guidelines, and suggestions for improvement. Future work on \softwarename\ can explore how AI might provide such feedback—not only by generating descriptions aligned with AD guidelines but also by offering targeted, guideline-based feedback on novice edits. This feedback can also be altered based on the content of the video described. For example, professional describers highlighted the research that goes into describing visual content. Hence, describers could benefit from AI assistance tailored to different video genres, such as educational or instructional, by providing content-specific information.
Additionally, given that AD guidelines vary across countries~\cite{rai2010comparative, bittner2012audio}, the system could also adapt its feedback to the describer’s region. However, relying solely on AI-generated evaluation may introduce issues like verbosity bias~\cite{saito2023verbosity}, where large language models (LLMs) tend to favor overly long descriptions compared to human preferences.
To mitigate this, future systems could incorporate direct feedback from BLV users, ensuring the descriptions align with user needs and preferences. By combining AI-driven suggestions with human-in-the-loop feedback, tools like \softwarename{} can better support novice describers in actively learning and refining their AD authoring skills.

Community-driven AD authoring tools can also empower novices to better integrate AI into their workflow. Participants frequently asked AI for help with aspects they were unfamiliar with or found challenging (e.g., timing, recognition of on-screen language). While professional describers often crafted targeted prompts for description refinement, novices frequently struggled to identify what needed improvement.
To bridge this gap, future work could introduce a library of commonly used prompts directly within the interface, helping novices learn how to engage with the AI more effectively and bringing their revision strategies closer to those of professionals.

\subsection{Flexible Control over AI Collaboration}
The collaborative AI agent should support the diverse workflows and stylistic preferences of AD authors. Professionals often edited both human- and AI-generated variations to more closely reflect their personal description style, and voiced concerns that AI could homogenize descriptions, potentially reducing the sense of immersion and overall experience for BLV users.
To address this, platforms that generate AI descriptions could allow users—especially professionals—to fine-tune models on their own authored content, with their consent. This level of personalization would reduce the number of manual edits needed while preserving the unique voice and aesthetic choices that are often critical in high-quality AD.

Our findings also reveal the importance of providing flexible, granular control over AI assistance in AD tools. While participants used AI to edit multiple descriptions, they preferred to edit descriptions manually for fine-grained changes. In addition, some participants expressed interest in generating descriptions using AI for only specific segments of a video. This segmented approach would allow describers to retain greater creative control while leveraging AI to describe labor-intensive or complex visuals.
The level of AI assistance needed also varies with experience. To accommodate different preferences and expertise levels, AD tools can offer customizable configurations, such as a ``guided mode'' for novices, with built-in AI assistance like structured feedback and suggestions, and a ``professional mode'' where the system learns from the describer’s editing patterns and prompt history. In addition, AI-based AD tools can allow BLV users to author descriptions through a ``BLV mode'' that offers non-visual editing affordances such as screen-reader support, Q\&A interactions, object identification, and AI prompting via conversational voice interfaces, enabling BLV users to actively participate in authoring descriptions. In this way, BLV users can actively contribute to AD creation as they bring their unique insights and perspectives to the AD process~\cite{jiang2023beyond, jiang2022co}. Bringing these different stakeholders together, such an AD tool can create an active ecosystem of AD training, collaboration, and personalization for video accessibility, offering a balance between AI support with human creativity and personalization in the AD authoring process.

 \subsection{Evolving AD Guidelines for Diverse Content}
 AD authoring tools offer a pathway to evolving general guidelines into nuanced, genre and format specific standards for diverse video content. The AD guidelines we used to create the descriptions did not focus on a specific video genre. Although these guidelines provide solid baseline support, when applied to specific video genres or video types, the descriptions might not match the intended viewing purpose. For instance, having too many details on the setting or environment for an instructional video can distract from the core instructional content. For novices, a lack of genre-specific guidelines could result in overlooking critical visual elements while including less relevant details. The challenges are further amplified for 360$^\circ$ videos, where there are even fewer guidelines, and prioritizing what to describe is challenging as their is no fixed point of view. 
Past work has shown that LLMs can effectively categorize videos into specific genres~\cite{li2025videoa11y}. \softwarename{} also enables users to input custom guidelines when generating AI descriptions. The combination of data on prompts, specific guidelines, and video categorization can help curate a new generation of adaptive AD guidelines, tailored to genre and video type. 

The change in AD practices is also being recognized at the regulatory level. Ofcom, the broadcast regulatory authority in the UK, has proposed changes to the existing AD guidelines, including the removal of requirements such as that ``delivery should be impersonal in style''~\cite{ofcom2023accessibility}. This signals a growing recognition of subjectivity and personalization in AD, making variations even more relevant. These variations can be useful for 360$^\circ$ video accessibility, where the 3D content experience is similar to 2D content with standard AD practices for BLV users~\cite{equalentry}. However, having multiple AD variations, each focusing on a different aspect of the 360 video (e.g., environment, action, spatial layout), can enrich the user experience without overwhelming them with audio descriptions. This strategy allows authors to collaboratively highlight complementary aspects of a scene, which is particularly useful in immersive and complex video formats. This reinforces that AD guidelines must be dynamic and responsive as new forms of video content emerge. Using a data-driven, community-informed approach based on user prompts, task types, and content categories to curate new guidelines and different variations would mark a significant shift away from static, one-size-fits-all guidelines, largely curated for film and television, and provide a scalable, sustainable framework for evolving AD practices.
 
 
 

\subsection{Social and Ethical Concerns of AI}
Generative AI has the potential to enhance individual creativity for tasks, but it also poses the risk of ideational convergence~\cite{doshi2024generative}. In \softwarename{}, we tried to mitigate these effects with features like variations and tags, which promote human collaboration and comparison. However, prior work suggests that BLV and sighted evaluators may focus on different aspects. BLV evaluators emphasize clarity and succinctness, while sighted evaluators focus more on descriptiveness and visual detail~\cite{natalie2021uncovering}. Building on this, future work can further improve collaborative aspects beyond editing (forking) by enabling BLV audiences to comment on and rate description variations, thereby supporting a broader diversity of styles tailored to end-user customization needs and preferences.

AI-assisted tools raise valid concerns about job displacement in creative fields such as audio description. However, \softwarename{} is designed to offload repetitive tasks, allowing users to focus on higher-level creative and editorial decisions. During recruitment, some professional describers were initially hesitant to participate in an AI-related project. To address their concerns, we provided a live walkthrough of the interface during the informed consent process. After hands-on experience, most participants responded positively, viewing the tool as a valuable complement to their existing workflow. By streamlining the authoring process, this approach also supports greater AD customization, enabling more variation and choice for BLV audiences. In doing so, this work contributes to a broader ethical conversation in accessibility research about how to harness rapid AI advances in partnership with human describers, while enhancing equitable access to audio description for BLV users.

From an environmental perspective, \softwarename{} uses GPT-4o, which requires significant computational resources and has an environmental impact due to energy consumption and water usage for cooling during inference~\cite{rillig2023risks}. We expect these environmental costs to evolve as new, more efficient models emerge. In future work, we plan to explore the use of smaller, task-specific models for lightweight operations such as grammatical and textual edits, while reserving larger models for complex, multimodal reasoning tasks like scene understanding. This hybrid approach may help reduce environmental impact while maintaining the quality and flexibility of the authoring experience.

\subsection{Limitations}
While our initial implementation of \softwarename{} demonstrates the potential of a collaborative AI-driven accessibility solution, it comes with several limitations.
First, our study involved 18 participants, split between novice and professional describers. This provided rich qualitative insights, but the generalizability of our quantitative findings remains limited due to the small sample size.
Second, participants only used \softwarename{} in a single hour session. Although most found the tool easy to use, long-term engagement would yield more meaningful insights into usability, interaction patterns, and the usefulness of collaborative features. This is particularly important for novice describers, many of whom were using an AD authoring tool for the first time and may need more time to develop confidence and proficiency. Longitudinal deployment of \softwarename{} is necessary to fully evaluate the usefulness and usability of these features.
Third, our system also included three human-generated variations per video. While this enabled comparison and supported learning, increasing the number of variations could lead to cognitive overload when selecting between them. Future work could explore mechanisms for filtering, sorting, or recommending variations to support more scalable use.
Fourth, we did not evaluate the revised descriptions with BLV users. Future studies will be essential to compare collaboratively edited descriptions with AI-generated ones and evaluate the AD variations in terms of quality, tone, and relevance for diverse BLV users. 
Finally, the videos used in the studies were limited in scope due to time constraints. Although Task 3 allowed participants to choose from four different videos, the selection did not capture the diversity of content found in real-world AD scenarios. Additionally, all tasks involved short-form videos, which may not reflect the complexity and narrative demands of longer-format media. A longitudinal, in-the-wild deployment of the tool would offer deeper insight into how describers adapt \softwarename{} to their workflows and a broader range of video content.

\section{Conclusion}
Our study underscores the value of \softwarename{} in advancing accessible video content through the integration of human expertise and AI support. By enabling a flexible, collaborative approach to audio description, the platform addresses key challenges faced by both novice and professional describers. Beyond enhancing individual workflows, \softwarename{} offers design insights for future tools that support and empower both accessibility trainees and professionals. In doing so, it paves the way for more inclusive and equitable media experiences, where diverse voices contribute to shaping accessible content for all.

\begin{acks}
 This research was supported by the National Eye Institute (NEI) of the National Institutes of Health (NIH) under award number R01EY034562. The content is solely the responsibility of the authors and does not necessarily represent the official views of the NIH. 
\end{acks}


\bibliographystyle{ACM-Reference-Format}
\bibliography{references}

\clearpage
\appendix

\section{Prompts for AI-Generated Descriptions}
\label{appendix:prompts}

\subsection{Generating Baseline AI Descriptions}
\label{appendix:video-prompt}

The following prompt is used to generate a baseline AI description for an input video:

\setlength{\fboxsep}{8pt} 
\setlength{\fboxrule}{0.5mm}
\vspace{0.2cm}
\noindent
\fcolorbox{black}{lightblue}{ 
    \begin{minipage}{0.94\linewidth}
    \raggedright
You are an AI designed to assist in creating high-quality and contextually rich descriptions for videos aimed at enhancing accessibility for blind and low-vision users. Your task is to generate video descriptions that are descriptive, objective, accurate, and clear while being fast and responsive in their creation. The descriptions should be highly personalized, based on interactive guidance from a human describer who will provide specific guidelines during the process. The input consists of a set of images representing key scenes or frames from the video. You must analyze these images to identify critical visual elements such as settings, characters, actions, emotions, and key objects. Using this analysis, you will construct a narrative that maintains logical continuity across scenes. The descriptions should ensure smooth transitions, providing the BLV user with a comprehensive and immersive understanding of the video's content. The goal is to improve accessibility, making the experience more engaging and inclusive. The descriptions should be customizable to different user preferences, offering an inclusive and personalized experience. If descriptions from previous API calls are available, they should be taken into account to ensure contextual relevance. \newline

\noindent\textbf{In the case where the user provided specific guidelines:}
\begin{quote}
Additionally, you should use both General Guidelines and Specific Guidelines when creating the descriptions. While both are important, the Specific Guidelines, which are defined by a human describer, should take priority and be given greater emphasis. For each description you generate, explicitly indicate which general guidelines were not followed and explain why they were ignored in the output. This will enhance explainability and transparency in your decision-making process. \newline \newline
\end{quote}
\textbf{Specific Guidelines:} \newline
\{User provided specific guidelines\} 

\textbf{General Guidelines:}
\begin{enumerate}
    \item Avoid over-describing and do not include non-essential visual details.
    \item Descriptions should not be opinionated unless the content demands it.
    \item Choose a level of detail based on plot relevance when describing scenes.
    \item Descriptions should be informative and conversational, using present tense and a third-person omniscient perspective.
    \end{enumerate}

    \end{minipage}
}

\setlength{\fboxsep}{8pt} 
\setlength{\fboxrule}{0.5mm}
\vspace{0.2cm}
\noindent
\fcolorbox{black}{lightblue}{ 
    \begin{minipage}{0.94\linewidth}
    \raggedright

    \begin{enumerate}
    \setcounter{enumi}{4}
    \item The vocabulary should reflect the predominant language or accent of the program and should be consistent with the genre and tone while being mindful of the target audience.
    \item Consider historical context and avoid words with negative connotations or bias.
    \item Use vivid verbs rather than bland ones with adverbs.
     \item Use pronouns only when it is clear whom they refer to.
    \item Use comparisons for shapes and sizes with familiar and globally relevant objects.
    \item Maintain consistency in word choice, character qualities, and visual elements across all audio descriptions.
    \item Ensure the tone and vocabulary match the target audience’s age range.
    \item Avoid errors in word selection, pronunciation, diction, or enunciation.
    \item Start with general context before adding details.
    \item Describe shape, size, texture, or color only when appropriate to the content.
    \item Use first-person narrative only when required to engage the audience.
    \item Use articles appropriately when introducing or referring to subjects.
    \item Prefer formal speech over colloquialisms unless appropriate for the content.
    \item When introducing new terms, objects, or actions, label them first and then follow with their definitions.
    \item Describe objectively without personal interpretation or commentary, and do not censor content.
    \item Deliver narration steadily and impersonally, but not monotonously, while matching the program's tone.
    \item Adjust the style for emotion and mood according to the program’s genre, adding excitement or lightness when appropriate.
    \item For children's content, tailor the language and pacing to suit their comprehension and feedback.
    \item Do not alter, filter, or exclude content—describe what you see while ensuring simplicity and succinctness.
    \item Prioritize relevance when describing actions to avoid affecting the user experience.
    \item Include location, time, and weather conditions if they are relevant to the scene or plot.
    \item Focus on key content for learning and enjoyment so that the intention of the program is effectively conveyed.
    \item When describing instructional content, present the sequence of activities first.
    \item For dramatic productions, highlight elements such as style, setting, focus, period, dress, facial features, objects, and aesthetics.
    \item Emphasize the most essential aspects of a scene to help the viewer follow, understand, and appreciate the content.
    \item Audio descriptions should include details about characters, locations, time, circumstances, on-screen actions, and on-screen text when relevant.

\end{enumerate}

    \end{minipage}
}

\setlength{\fboxsep}{8pt} 
\setlength{\fboxrule}{0.5mm}
\vspace{0.2cm}
\noindent
\fcolorbox{black}{lightblue}{ 
    \begin{minipage}{0.94\linewidth}
    \raggedright
    \begin{enumerate}
    \setcounter{enumi}{30}
    \item Describe only what a sighted viewer can see.
    \item When describing characters, prioritize factual traits such as hair, skin, eye color, build, height, age, and visible disabilities while ensuring consistency. Use person-first language and avoid singling out characters for specific traits unless they are relevant to the story.
    \item If racial, ethnic, or gender identity is not confirmed or established in the plot, do not make assumptions.
    \item When introducing characters for the first time, aim to include a descriptor before the name (e.g., "a bearded man, Jack").
    \item Descriptions should convey facial expressions, body language, and reactions.
    \item If race is important to the meaning or intent of the content, describe it using currently accepted terminology.
    \item Avoid identifying characters solely by gender expression unless it provides unique insights not otherwise apparent to visually impaired viewers.
    \item Describe character clothing if it enhances characterization, plot, setting, or genre enjoyment.
    \item If on-screen text is central to understanding, establish a pattern of reading the words aloud and announce when text appears.
    \item In the case of subtitles, read the translation after stating that a subtitle appears.
    \item When shot changes are crucial to understanding a scene, indicate them by describing where the action takes place or where characters are positioned in the new shot.
    \item Provide descriptions before the content rather than after.
    Keep descriptions between 25 to 50 words in length.
    \end{enumerate}
    \end{minipage}
}

\subsection{Generating AD Tags}
\label{sec:customize}

The following prompt is used to generate tags for the videos:

\setlength{\fboxsep}{8pt} 
\setlength{\fboxrule}{0.5mm}
\vspace{0.2cm}
\noindent
\fcolorbox{black}{lightblue}{ 
    \begin{minipage}{0.94\linewidth}
    \raggedright
You are an AI designed to assist in analyzing the provided list of descriptions and identifying a set of up to four most relevant keywords from the following categories. Your response should consist solely of a Python list of keywords, with no further explanations, formatting, or extra text.
\label{appendix:tags-prompt}
\begin{itemize}
    \item \textbf{Description Length}: [Concise, Complete description]
    \item \textbf{Focus}: [Main story focus, Character focus, Environment focus]
    \item \textbf{Interpretations}: [With Interpretations, Without Interpretations]
    \item \textbf{Detail Level}: [Low detail, Medium detail, High detail]
    \item \textbf{Action Description}: [Detailed action, Brief action, No action]
    \item \textbf{Character Details}: [Character-driven, Action-driven, Environmental focus]
    \item \textbf{Tagging of Key Visuals/Objects}: [Key visuals highlighted, Important objects tagged, Minimal object tagging]
\end{itemize}



    \end{minipage}
}

\setlength{\fboxsep}{8pt} 
\setlength{\fboxrule}{0.5mm}
\vspace{0.2cm}
\noindent
\fcolorbox{black}{lightblue}{ 
    \begin{minipage}{0.94\linewidth}
    \raggedright
\label{appendix:tags-prompt}
\begin{itemize}
    \item \textbf{Environmental Description}: [Detailed environment, Basic environment, Environment-free]
\end{itemize}

\noindent You can select a maximum of one keyword from each category, for a total of up to four keywords. Additionally, feel free to include up to two additional keywords that may not be listed, if relevant.
You should generate two lists. One list includes the keywords and the other one includes additional\_keywords. \newline

\noindent Descriptions: \newline
\{AI generated initial descriptions\}

    \end{minipage}
}

\subsection{Revising Descriptions with AI Prompting}
\label{appendix:input-prompt-prompt}

The following prompt is used to generate revised descriptions based on a describer's input prompt:

\setlength{\fboxsep}{8pt} 
\setlength{\fboxrule}{0.5mm}
\vspace{0.2cm}
\noindent
\fcolorbox{black}{lightblue}{ 
    \begin{minipage}{0.94\linewidth}
    \raggedright
You are an advanced AI designed to enhance and refine audio descriptions for videos, specifically aimed at improving accessibility for blind and low-vision (BLV) users. Your task is to customize these descriptions based on input prompt, ensuring they are tailored to the specific needs and preferences of the user. \newline
Your role is to revise the provided descriptions, making necessary adjustments to align with the user's goals—whether it's increasing detail, adjusting the tone, improving clarity or something else. Your objective is to ensure that each description not only meets the prompt's requirements but also enhances the overall accessibility and experience for BLV users. \newline \newline
\textbf{Input Prompt:} \newline
\{User provided prompt\} \newline 
\textbf{Description:} \newline
\{User provided description to be revised\} \newline \newline
You will also be provided with relevant video frames. Use them to enrich the descriptions where necessary.
**Return only the revised description with no introductory text, explanations, or additional commentary.**

    \end{minipage}
}

\subsection{Codebook of Prompt Types with Definitions and Examples}
\label{appendix:codebook}
\newpage 
\begin{table*}[ht]
\centering
\begin{tabular}{p{2cm}p{7cm}p{6cm}}
\toprule
\textbf{Code Name} & \textbf{Code Description} & \textbf{Prompt Example} \\
\toprule
Simplify & Requests to make the description easier to understand, or remove unnecessary details. &``Make it more accessible for ESL users.'' \\
Shorten & Requests to make the description shorter. & ``Condense, use shorter syllables.'' \\
Remove & Requests to remove specific elements (e.g., objects, emotions). & ``Remove the repetitive background setting description for the second description.'' \\
Replace & Requests to replace a word or phrase with another. & ``Change `a woman' to `our guide.'~'' \\
Addition & Requests to add specific visual details. & ``Add a little bit more detail before the boy fallen asleep.'' \\
Text on screen & Requests to add specific text on screen to description. & ``please transcribe the text on screen'' \\
Correction & Comments by the user based on any hallucination in the description. & ``There are no colorful blinders.'' \\
Description Flow & Requests related to order, or structuring text differently. & ``Describe the room first then the boy.'' \\
Language & Requests related to translating or structuring sentences based on language requirements (e.g., active voice, emotional descriptors). & ``Make passive to active verbs''\\
Reduce Detail & Requests to reduce unnecessary detail on objects. & ``Less detail on mixing bowl'' \\
Query & Questions asked by describers on elements in the video. & ``Can you tell me what's the language on screen that isn't English?'' \\
Miscellaneous & Requests that cannot be categorized in others (e.g., requests to edit timestamps, add new descriptions). & ``Help me combine these two description as one.'' \\
\bottomrule
\end{tabular}
\end{table*}

\end{document}